\documentclass[prd,twocolumn,notitlepage,eqsecnum,nofootinbib,unsortedaddress]{revtex4-1}
\usepackage{amssymb}
\usepackage{amsmath}
\usepackage{mathdots}
\usepackage{bm}
\usepackage{color}
\usepackage[normalem]{ulem}
\begin{document}
\allowdisplaybreaks
\title{Regularization of static self-forces}   
\author{Marc Casals} 
\affiliation{Perimeter Institute for Theoretical Physics, Waterloo,
  Ontario, Canada N2L 2Y5} 
\affiliation{Department of Physics, University of Guelph, Guelph,
  Ontario, Canada N1G 2W1} 
\affiliation{School of Mathematical Sciences and Complex \& Adaptive
  Systems Laboratory, University College Dublin, Belfield, Dublin 4, Ireland} 
\author{Eric Poisson} 
\author{Ian Vega} 
\affiliation{Department of Physics, University of Guelph, Guelph,
  Ontario, Canada N1G 2W1} 
\date{June 17, 2012} 
\begin{abstract} 
Various regularization methods have been used to compute the
self-force acting on a static particle in a static, curved
spacetime. Many of these are based on Hadamard's two-point function in 
three dimensions. On the other hand, the regularization method that
enjoys the best justification is that of Detweiler and Whiting, which
is based on a four-dimensional Green's function. We establish the
connection between these methods and find that they are all
equivalent, in the sense that they all lead to the same static
self-force. For general static spacetimes, we compute local expansions
of the Green's functions on which the various regularization methods
are based. We find that these agree up to a certain high order, and
conjecture that they might be equal to all orders. We show that this
equivalence is exact in the case of ultrastatic spacetimes. Finally,
our computations are exploited to provide regularization parameters
for a static particle in a general static and spherically-symmetric
spacetime. 
\end{abstract} 
\maketitle

\section{Introduction} 

A test body moving freely in a curved spacetime follows a geodesic
of the spacetime. When, however, the body carries a (scalar or
electric) charge, the field created by the charge interacts with the
spacetime curvature in such a way as to produce a deformation of 
the field lines from an otherwise isotropic distribution around the
body. The field gives rise to a net self-force acting on the body,
and the self-force prevents it from moving on a geodesic. The
self-force typically contains two components, a radiation-reaction
force that is accompanied by a loss of energy to radiation, and a
conservative force that survives even when the body is maintained in a
stationary position. A self-force can also be present in the absence
of a charge, when the body's mass is too large for it to be considered
a test mass; in this case the body creates a gravitational perturbation 
that affect its motion, which is no longer geodesic in the background
spacetime. The (scalar, electromagnetic, and gravitational) self-force
has been the topic of intense development in the last several years;
for an extensive review see Ref.~\cite{poisson-pound-vega:11}. Most of
this activity was focused on the gravitational case, in an effort to
model the inspiral and gravitational-wave emissions of a binary system
with a small mass ratio \cite{barack-sago:10, diener-etal:11,
  warburton-etal:12}.  

Self-force computations are usually attempted under the assumption
that the body is a point particle, in order to avoid the largely
irrelevant complications associated with internal structure. In this
context, however, the very definition of the self-force requires
scrutiny. Given that the field of a point particle diverges at the
position occupied by the particle, it is not immediately clear how one
can make sense of its action on the particle and construct a
self-force that is well defined, finite, and in agreement with the
self-force acting on an extended body in the the limit in which the
size is taken to zero. One must find a sensible regularization
procedure that not only returns a finite expression for the
self-force, but does so in a unique and physically well-motivated  
way. In this paper we examine regularization procedures that have been
invoked in the computation of (scalar and electromagnetic) self-forces
in the restricted context of {\it static particles in static
  spacetimes}. Our aim is to show that the differing procedures are
equivalent and lead to the same self-force. To the best of our
knowledge, this issue has not been previously addressed in the
literature.  

In our view, the regularization procedure that has received the best
physical and mathematical justification is the one proposed by
Detweiler and Whiting \cite{detweiler-whiting:03}. The method, which
is completely general and not restricted to static situations,
involves a decomposition of the field created by the particle into
singular and regular pieces. The singular field is precisely
identified by a local construction, and is designed to provide an
exact solution to the field equation sourced by the particle, with the
property that it shares the singularity structure of the particle's
actual field. The regular field is the difference between the actual
field and the singular field; it satisfies the source-free field
equation, it is smooth at the particle's position, and it is known to
be entirely responsible for the self-force. The Detweiler-Whiting
regularization method has been thoroughly justified \cite{harte:06,
  harte:09, pound:10a, harte:11}, and it has emerged as the method of
choice in most self-force computations reported in the recent
literature. Because of its generality and naturalness, it is the
standard by which other regularization methods must be compared.  

Many self-force computations, however, did not make use of the
Detweiler-Whiting regularization procedure, but employed instead 
{\it ad hoc} procedures that perhaps do not enjoy the same degree of
justification. This is the case of all computations of self-forces
acting on static particles in static spacetimes \cite{smith-will:80,
  zelnikov-frolov:82, linet:86, boisseau-charmousis-linet:96,
  wiseman:00, linet:00, shankar-whiting:07, linet:07,
  khusnutdinov-bakhmatov-ilya:07, bezerra-etal:09,
  khusnutdinov-etal:10, bezerra-saharian:07,
  barbosa-spinelly-bezerra:09, barbosa-freitas-bezerra:10, 
  bezerra-saharian:12}, which involved a variety of regularization    
methods. In the pioneering Smith-Will paper \cite{smith-will:80}, for 
example, the field of a static electric charge in the spacetime of a
Schwarzschild black hole was regularized by the Copson solution
\cite{copson:28}, which was shown to be as singular as the particle's
own field but to exert no force. As other examples, self-force
computations for charges in wormhole spacetimes \cite{linet:07,
  khusnutdinov-bakhmatov-ilya:07, bezerra-etal:09,
  khusnutdinov-etal:10}, or for charges near global 
monopoles \cite{bezerra-saharian:07,
  barbosa-spinelly-bezerra:09, barbosa-freitas-bezerra:10, 
  bezerra-saharian:12}, were regularized with the help of Hadamard's  
two-point function, defined in each spatial section of the
four-dimensional spacetime (or in a conformally related
space). Because Copson's solution is known to be an exact
representation of Hadamard's function in the (conformally related)
spatial sections of the Schwarzschild spacetime, these regularization
methods are essentially the same.  

The issue that interests us in this paper is the relationship between
these regularization procedures, and whether they can be shown to be 
equivalent, so that they will lead to the same self-force. The 
regularization procedures mentioned previously are all based on a
choice of Green's function for the (scalar or electromagnetic)
field. We shall consider a number of possible choices. 

The first is the four-dimensional version of the Detweiler-Whiting
singular Green's function, given by 
\begin{equation} 
G^{\sf S}_4(x,x') = \frac{1}{2} U(x,x') \delta(\sigma) 
- \frac{1}{2} V(x,x') \Theta (\sigma), 
\label{DW_def}
\end{equation} 
in which $x$ and $x'$ are spacetime events, assumed to be sufficiently
close that they are within each other's normal convex neighborhood, 
$\sigma := \sigma(x,x')$ is Synge's world function, equal to half the
squared geodetic distance between $x$ and $x'$, $\Theta$ is the
Heaviside step function, $\delta$ is the Dirac distribution, and $U$,
$V$ are two-point functions that are known to be smooth when 
$x \to x'$. Because $\sigma = 0$ when $x$ and $x'$ are linked by a
null geodesic, we see that the Green's function is singular
on the past and future light cones emerging from $x'$, and has support
outside the light cones, where $\sigma > 0$; it is also symmetric in
its arguments. As stated previously, the Detweiler-Whiting Green's
function gives rise to a robust regularization procedure that applies
to any particle moving in any spacetime.    

For static particles in static spacetimes, an adequate substitute for
the four-dimensional singular Green's function is its
three-dimensional variant 
\begin{equation} 
G^{\sf S}_3(\bm{x},\bm{x'}) = \int G^{\sf S}_4(x,x')\, d\tau',  
\end{equation} 
obtained by integrating the four-dimensional Green's function over 
the proper time $\tau'$ of a static observer at the spatial position
$\bm{x'}$. For static particles in static spacetimes, this Green's
function gives rise to the same regularization procedure as the
four-dimensional version. 

An alternative choice of Green's function, which is also appropriate
in the case of static particles in static spacetimes, is the
three-dimensional Hadamard function, given by 
\begin{equation} 
G^{\sf H}_3(\bm{x},\bm{x'}) 
= \frac{W(\bm{x},\bm{x'})}{\sqrt{2\sigma}}, 
\label{H_def} 
\end{equation} 
in which $\sigma(\bm{x},\bm{x'})$ is now half the squared geodetic
distance between $\bm{x}$ and $\bm{x'}$ as measured in the purely 
spatial sections of the spacetime, and $W$ is a smooth two-point
function. This Green's function, also known as Hadamard's elementary
solution \cite{hadamard:23}, is a local construction that is known to
reproduce the singular behavior of a field sourced by a point particle
at $\bm{x'}$. It is a plausible starting point for a regularization
procedure, but as stated above, it does not enjoy the same level of
justification as the Detweiler-Whiting Green's function. 

As a final choice we shall also consider a Green's function
$\tilde{G}^{\sf H}_3(\bm{x},\bm{x'})$ that is related to the Hadamard
function by a conformal transformation. This is to account for the
fact that it is often convenient, when solving for the field in the
spatial sections of the spacetime, to formulate the field equation in
a conformally related space. If the metric on the original spatial
sections is $h_{ab}$, then the metric on the conformally related space 
is $\tilde{h}_{ab} = \Omega^{-2} h_{ab}$, in which $\Omega(\bm{x})$ is
a scalar field. When $\tilde{h}_{ab}$ is simple, the field equation
simplifies in the conformally related space, and the field can then be
regularized with the help of $\tilde{G}^{\sf H}_3$, which differs from
$G^{\sf H}_3$ by factors of $\Omega$. 

Our main goal in this paper is to compare the regularization
procedures that are based on $G^{\sf S}_3(\bm{x},\bm{x'})$, 
$G^{\sf H}_3(\bm{x},\bm{x'})$, and 
$\tilde{G}^{\sf H}_3(\bm{x},\bm{x'})$. Our preferred regularization 
method is the one based on the Detweiler-Whiting Green's function,
because the resulting singular field was proved to share the same 
singularity structure as the particle's actual field and to exert no
force on the particle. In this case we find that if $\epsilon$ is a
measure of distance between $\bm{x}$ and $\bm{x'}$, then the singular
Green's function admits the local expansion  
\begin{equation} 
G^{\sf S}_3(\bm{x},\bm{x'}) = \frac{1}{\epsilon} \Bigl[ 1 +   
g_1 \epsilon + g_2 \epsilon^2 + g_3 \epsilon^3 + O(\epsilon^4)
\Bigr], 
\end{equation} 
with expansion coefficients $g_1$, $g_2$, and $g_3$ (computed below)
that depend on geometrical quantities (such as the spatial Riemann
tensor) evaluated at $\bm{x'}$. Removing this from the particle's
actual field returns a regularized field that is finite (indeed, once  
differentiable) at the position of the particle, leading to a
straightforward computation of the self-force.  

Our main result is the statement that the local expansions of 
$G^{\sf H}_3(\bm{x},\bm{x'})$ and 
$\tilde{G}^{\sf H}_3(\bm{x},\bm{x'})$ are identical to the local
expansion of the Detweiler-Whiting Green's function, and that they
therefore lead to equivalent regularization procedures. We view this
as a significant result that clarifies and justifies the alternative
regularization methods that have been employed in self-force
computations; it puts these computations on a firmer footing, and
lends them additional credence. 

We show by explicit calculation that the local expansions of all 
Green's functions agree through order $\epsilon^2$, which is more than
enough to guarantee the same self-force, but we also ask
whether equality could be established to all orders in $\epsilon$. We
provide only a partial answer to this question. We collect evidence
that the conjectured equality is likely to be true, sketch a proof
that relies on a strong assumption about the convergence of formal 
power series, and present a complete proof of equality in the special
case of ultrastatic spacetimes. 

As an application of our results we consider the scalar and
electromagnetic self-force acting on a particle held in place in a
static and spherically-symmetric spacetime. The assumption of
spherical symmetry implies that the self-force can be easily computed
with a mode-sum method based on a spherical-harmonic decomposition of  
the field. The mode coefficients of the singular field then play the
role of {\it regularization parameters} \cite{barack-ori:00,
  barack-etal:02, barack-ori:03a} that can be inserted in the mode-sum
to ensure convergence. We use our local expansions to compute these
regularization parameters for any static and spherically-symmetric
spacetime. We expect that our explicit listing of regularization
parameters will greatly facilitate future self-force computations.    

Our developments below rely heavily on the general theory of bitensors
and Green's functions in curved spacetime, as developed in
Ref.~\cite{dewitt-brehme:60} and summarized in
Ref.~\cite{poisson-pound-vega:11}, from which we import our 
notations. We begin in Sec.~\ref{sec:static} with a description of the
static spacetimes that are implicated in this work. We continue in
Sec.~\ref{sec:field_green} with a review of the scalar and
electromagnetic field equations in static spacetimes, along with the
associated Green's functions, and in Sec.~\ref{sec:Hadamard} we review
the Hadamard construction of the three-dimensional Green's
functions. The following sections contain our new work. In
Secs.~\ref{sec:local} and \ref{sec:local_conformal} we compute the
local expansions of the Hadamard functions. We do the same for the
Detweiler-Whiting functions in Sec.~\ref{sec:DW}, and prove the 
equality of the local expansions in Sec.~\ref{sec:equality}. In
Sec.~\ref{sec:equality_ultrastatic} we prove that in ultrastatic
spacetimes, the Hadamard and Detweiler-Whiting functions are 
strictly equal to one another. And finally, in Sec.~\ref{sec:spherical}
we consider the scalar and electromagnetic self-forces in static,
spherically-symmetric spacetimes, and involve the local expansions in
a computation of regularization parameters for mode-sum computations
of the self-force. 

In the following we denote a spacetime event by $x$ or $x'$, and a 
spatial position by $\bm{x}$ or $\bm{x'}$, so that $x = (t,\bm{x})$
and $x' = (t',\bm{x'})$. Spacetime tensors at $x$ are denoted
$A^\alpha$, with a Greek index $\alpha$ that ranges over the values
$\{0, 1, 2,3\}$; spacetime tensors at $x'$ are denoted $A^{\alpha'}$,
with a primed Greek index. Spatial tensors at $\bm{x}$ are denoted
$A^a$, with a Latin index $a$ that ranges over $\{1,2,3\}$; spatial
tensors at $\bm{x'}$ are denoted $A^{a'}$, with a primed Latin
index. Variants of these notations will be introduced as needed.   

\section{Static spacetimes} 
\label{sec:static} 

The class of spacetimes considered in this paper admits a
hypersurface-orthogonal, timelike Killing vector $t^\alpha$, and in an
adapted coordinate system the metric is expressed as 
\begin{equation} 
ds^2 = -N^2\, dt^2 + h_{ab}\, dx^a dx^b, 
\label{metric} 
\end{equation} 
in terms of a lapse function $N$ and a spatial metric $h_{ab}$ which
depend on the spatial coordinates $x^a$ only. No other assumptions are
placed on the spacetime. We introduce the vector field 
\begin{equation} 
A_a := \partial_a \ln N = \frac{\partial_a N}{N}, 
\label{A_def}
\end{equation} 
which acts as a substitute for $\partial_a N$; the vector has a
vanishing time component. 

A straightforward computation reveals that the connection coefficients
are given by 
\begin{equation} 
\mbox{}^4 \Gamma^t_{t a} = A_a, \qquad 
\mbox{}^4 \Gamma^a_{tt} = N^2 A^a, \qquad 
\mbox{}^4 \Gamma^a_{bc} = \Gamma^a_{bc}, 
\label{connection} 
\end{equation} 
in which $\Gamma^a_{bc}$ is the connection compatible with the spatial
metric $h_{ab}$. We shall indicate covariant differentiation relative
to the spacetime connection with the operator $\nabla_\mu$ or with  
a semicolon (for example, $\nabla_\mu A^\nu = A^\nu_{\ ;\mu}$), and
covariant differentiation relative to the spatial connection with the
operator $D_a$ or with a vertical stroke (for example,
$D_a A^b = A^b_{\, |a}$).  

For future reference we examine a vector $v^\alpha$ that is
parallel-transported along a purely spatial curve described by the
parametric equations $t = \mbox{constant}$, $x^a = z^a(s)$, in which
$s$ is proper distance. The vector tangent to the curve is $n^\alpha$
with components $n^t = 0$ and $n^a = dz^a/ds$. The equation of
parallel transport is 
\begin{equation} 
\frac{dv^\alpha}{ds} + \mbox{}^4 \Gamma^\alpha_{\beta\gamma} 
v^\beta n^\gamma = 0, 
\end{equation} 
and its time component reduces to $dv^t/ds + (A_a n^a) v^t = 0$. With
$A_a n^a = N^{-1} dN/ds$, the solution to the differential equation is
$N v^t = \mbox{constant}$, or 
\begin{equation} 
v^t(s) = \frac{N(0)}{N(s)}\, v^t(0). 
\label{parallel_transport} 
\end{equation} 
For the spatial components we find that the equation reduces to
$dv^a/ds + \Gamma^a_{bc} v^b n^c = 0$, which states that $v^a$ is
parallel-transported as if it were a vector in a three-dimensional
space with metric $h_{ab}$. When the spatial curve is a spacetime
geodesic, we find that Eq.~(\ref{parallel_transport}) produces 
$n^t(s) = 0$, while the spatial components reveal that $n^a$
satisfies the geodesic equation in the three-dimensional space. A
purely spatial geodesic in spacetime is therefore a geodesic in a 
three-dimensional space with metric $h_{ab}$.  

The nonvanishing components of the Riemann tensor are 
\begin{equation} 
\mbox{}^4 R_{tatb} = N^2 \bigl( A_{a|b} + A_a A_b \bigr), \qquad 
\mbox{}^4 R_{abcd} = R_{abcd}, 
\label{Riemann} 
\end{equation} 
in which $R_{abcd}$ is the Riemann tensor associated with the spatial
metric $h_{ab}$. The nonvanishing components of the Ricci tensor are 
\begin{subequations} 
\label{Ricci_tensor}
\begin{align}
\mbox{}^4 R_{tt} &= N^2 \bigl( A^c_{\ |c} + A^c A_c \bigr), \\
\mbox{}^4 R_{ab} &= R_{ab} - \bigl( A_{a|b} + A_{a} A_{b} \bigr). 
\end{align}
\end{subequations}  
The Ricci scalar is 
\begin{equation} 
\mbox{}^4 R = R - 2 \bigl( A^c_{\ |c} + A^c A_c \bigr). 
\label{Ricci_scalar} 
\end{equation} 
For future reference we also record the components of the covariant
derivative of the Riemann tensor: 
\begin{subequations}
\label{Riemann_cov} 
\begin{align} 
\mbox{}^4 R_{tatb;c} &= N^2 \bigl( A_{a|bc} + A_{a|c} A_b 
+ A_a A_{b|c}  \bigr), \\ 
\mbox{}^4 R_{tabc;t} &= N^2 \bigl( A_{a|b} A_c - A_{a|c} A_b 
- A^d R_{dabc} \bigr), \\ 
\mbox{}^4 R_{abcd;e} &= R_{abcd|e}. 
\end{align} 
\end{subequations} 
We also have 
\begin{subequations} 
\label{Riccit_cov} 
\begin{align} 
\mbox{}^4 R_{tt;a} &= N^2 \bigl( A^c_{\ |ca} + 2 A^c A_{c|a} \bigr), 
\\ 
\mbox{}^4 R_{ta;t} &= -N^2 \bigl( A^c_{\ |c} A_a - A^c_{\ |a} A_c 
+ A^c R_{ca} \bigr),  \\ 
\mbox{}^4 R_{ab;c} &= R_{ab|c} - A_{a|bc} - A_{a|c} A_b - A_a A_{b|c} 
\end{align} 
\end{subequations} 
and 
\begin{equation} 
\mbox{}^4 R_{;a} = R_{|a} -2 \bigl( A^c_{\ |ca} + 2 A^c A_{c|a} \bigr) 
\label{Riccis_cov}
\end{equation} 
for the covariant derivative of the Ricci tensor and scalar. 

Below we shall consider a scalar or electric charge at rest in the
static spacetime. The charge follows an orbit of the timelike Killing
vector, and the only nonvanishing component of its velocity vector is 
\begin{equation} 
u^t = \frac{1}{N}. 
\label{u} 
\end{equation} 
The covariant acceleration is defined by 
$a^\alpha := u^\alpha_{\ ;\beta} u^\beta$, and its nonvanishing
components are 
\begin{equation} 
a^a = A^a.
\label{a}
\end{equation} 
We shall also require the vectors $\dot{a}^\alpha := 
a^\alpha_{\ ;\beta} u^\beta$ and $\ddot{a}^\alpha := 
\dot{a}^\alpha_{\ ;\beta} u^\beta$; the nonvanishing components are 
\begin{equation} 
\dot{a}^t = \frac{1}{N} A^c A_c, \qquad 
\ddot{a}^a = \bigl( A^c A_c \bigr) A^a. 
\label{adot} 
\end{equation} 

\section{Field equations and Green's functions} 
\label{sec:field_green} 

\subsection{Scalar field} 
\label{subsec:fg_scalar} 

The potential $\Phi$ generated by a scalar-charge density $\mu$
obeys the wave equation 
\begin{equation} 
\Box \Phi = -4\pi \mu 
\label{Phi_wave} 
\end{equation} 
in any four-dimensional spacetime; 
$\Box := g^{\alpha\beta} \nabla_\alpha \nabla_\beta$ is the covariant
wave operator. Our considerations in this paper are limited to scalar
fields that are minimally coupled to the spacetime curvature; it is,
however, a very straightforward exercise to extend our discussion
to arbitrary couplings. When the spacetime is static, and when the
potential and charge density are both time-independent, the wave
equation reduces to  
\begin{equation} 
\nabla^2 \Phi + A^a \partial_a \Phi = -4\pi \mu;  
\label{Phi_Poisson} 
\end{equation} 
here $\nabla^2 := h^{a b} D_a D_b$ is the covariant Laplacian operator
in a three-dimensional space with metric $h_{ab}$. 

It is sometimes convenient to formulate Eq.~(\ref{Phi_Poisson}) in 
a conformally related space with metric $\tilde{h}_{ab}$; this is
related to the original metric $h_{ab}$ by the conformal
transformation   
\begin{equation} 
h_{ab} = \Omega^2 \tilde{h}_{ab}, 
\label{conf_trans} 
\end{equation} 
in which $\Omega$ is a function of the spatial coordinates $x^a$. As a
consequence of this transformation we find that $h^{ab} = \Omega^{-2}
\tilde{h}^{ab}$ and $h^{1/2} = \Omega^3 \tilde{h}^{1/2}$. A simple
computation reveals that in the conformal formulation,
Eq.~(\ref{Phi_Poisson}) becomes  
\begin{equation} 
\tilde{\nabla}^2 \Phi + \tilde{A}^a \partial_a \Phi 
= -4\pi \tilde{\mu}, 
\label{Phi_P_conf} 
\end{equation} 
in which $\tilde{\nabla}^2 := \tilde{h}^{ab} \tilde{D}_a \tilde{D}_b$
is the covariant Laplacian operator associated with the metric
$\tilde{h}_{ab}$, 
\begin{equation} 
\tilde{A}^a := \tilde{h}^{ab} \partial_b \ln ( N\Omega) 
= \tilde{h}^{ab} \biggl( \frac{\partial_b N}{N} 
+ \frac{\partial_b \Omega}{\Omega} \biggr), 
\label{A_tilde} 
\end{equation} 
and 
\begin{equation} 
\tilde{\mu} := \Omega^2 \mu. 
\label{mu_tilde} 
\end{equation} 

The four-dimensional Green's function associated with
Eq.~(\ref{Phi_wave}) is $G_4(x,x')$, which satisfies the wave equation  
\begin{equation} 
\Box G_4(x,x') = -4\pi \delta_4(x,x'), 
\label{G4_wave} 
\end{equation} 
in which $\delta_4(x,x')$ is a scalarized Dirac distribution defined
by 
\begin{equation} 
\delta_4(x,x') = \frac{\delta(x-x')}{\sqrt{-g'}}, 
\label{delta4}
\end{equation} 
where $\delta(x-x')$ is the usual product of four coordinate
delta functions. A solution to Eq.~(\ref{Phi_wave}) is then given by 
\begin{equation} 
\Phi(x) = \int G_4(x,x') \mu(x')\, \sqrt{-g'} d^4x'. 
\label{Phi_wavesol} 
\end{equation} 

The three-dimensional Green's function associated with
Eq.~(\ref{Phi_Poisson})  is $G_3(\bm{x},\bm{x'})$, which satisfies the 
Poisson equation 
\begin{equation} 
\nabla^2 G_3(\bm{x},\bm{x'}) 
+ A^a \partial_a G_3(\bm{x},\bm{x'}) = -4\pi \delta_3(\bm{x},\bm{x'}), 
\label{G3_Poisson} 
\end{equation} 
in which $\delta_3(\bm{x},\bm{x'})$ is a scalarized Dirac distribution
defined by 
\begin{equation} 
\delta_3(\bm{x},\bm{x'}) = \frac{\delta(\bm{x}-\bm{x'})}{\sqrt{h'}}, 
\label{delta3} 
\end{equation} 
where $\delta(\bm{x}-\bm{x'})$ is the usual product of three
coordinate delta functions.  A solution to Eq.~(\ref{Phi_Poisson}) is
then given by 
\begin{equation} 
\Phi(\bm{x}) = \int G_3(\bm{x},\bm{x'}) \mu(\bm{x'})\, 
\sqrt{h'} d^3x'.  
\label{Phi_Poisson_sol} 
\end{equation} 

The Green's function associated with Eq.~(\ref{Phi_P_conf})  is 
$\tilde{G}_3(\bm{x},\bm{x'})$, which satisfies  
\begin{equation} 
\tilde{\nabla}^2 \tilde{G}_3(\bm{x},\bm{x'}) 
+ \tilde{A}^a \partial_a \tilde{G}_3(\bm{x},\bm{x'}) 
= -4\pi \tilde{\delta}_3(\bm{x},\bm{x'}), 
\label{G3_P_conf} 
\end{equation} 
in which $\tilde{\delta}_3(\bm{x},\bm{x'})$ is defined by 
\begin{equation} 
\tilde{\delta}_3(\bm{x},\bm{x'}) =
\frac{\delta(\bm{x}-\bm{x'})}{\sqrt{\tilde{h}'}}.  
\label{delta3_conf} 
\end{equation} 
A solution to Eq.~(\ref{Phi_P_conf}) is then given by 
\begin{equation} 
\Phi(\bm{x}) = \int \tilde{G}_3(\bm{x},\bm{x'}) \tilde{\mu}(\bm{x'})\,  
\sqrt{\tilde{h}'} d^3x'.   
\label{Phi_Psol_conf} 
\end{equation} 

The relation between $G_4$ and $G_3$ can be identified by performing
the time integration in Eq.~(\ref{Phi_wavesol}) and comparing with
Eq.~(\ref{Phi_Poisson_sol}). The result is  
\begin{equation} 
G_3(\bm{x},\bm{x'}) = \int G_4(x,x')\, N(\bm{x'}) dt'. 
\label{G3_and_G4}
\end{equation} 
Apart from the factor of $N(\bm{x'})$, which converts from coordinate 
time to proper time at $\bm{x'}$, the three-dimensional Green's
function is simply the time integral of the four-dimensional Green's 
function. 

The relation between $\tilde{G}_3$ and $G_3$ is found by making the
substitutions $h^{1/2} = \Omega^3 \tilde{h}^{1/2}$ and $\mu =
\Omega^{-2} \tilde{\mu}$ within Eq.~(\ref{Phi_Poisson_sol}) and
comparing with Eq.~(\ref{Phi_Psol_conf}). The result is 
\begin{equation} 
\tilde{G}_3(\bm{x},\bm{x'}) = \Omega(\bm{x'}) G_3(\bm{x},\bm{x'}).  
\label{G3tilde_vs_G3} 
\end{equation}   
This can be confirmed by  expressing Eq.~(\ref{G3_Poisson}) in terms
of the conformally-related metric $\tilde{h}_{ab}$. We find that the
equation becomes     
\begin{equation} 
\tilde{\nabla}^2 G_3(\bm{x},\bm{x'})  
+ \tilde{A}^a \partial_a G_3(\bm{x},\bm{x'}) 
= -4\pi \frac{\delta(\bm{x}-\bm{x'})}
                     {\Omega' \sqrt{\tilde{h}'}}, 
\end{equation} 
and comparison with Eq.~(\ref{G3_P_conf}) allows us to make the
identification of Eq.~(\ref{G3tilde_vs_G3}). 
   
When the charge density $\mu$ describes a point charge $q$ moving on a
world line $\gamma$ described by the parametric relations $z(\tau)$,
we have that  
\begin{equation} 
\mu(x) = q \int_\gamma \delta_4\bigl( x, z(\tau) \bigr)\, d\tau. 
\label{mu_point} 
\end{equation} 
For a general world line the scalar charge produces a potential given
by Eq.~(\ref{Phi_wavesol}), which evaluates to 
\begin{equation} 
\Phi(x) = q \int_\gamma G_4\bigl(x,z(\tau)\bigr)\, d\tau. 
\label{Phi_gamma} 
\end{equation}  
For a static charge at a fixed position $\bm{z}$, the integral of
Eq.~(\ref{mu_point}) evaluates to 
\begin{equation} 
\mu(\bm{x}) = q \delta_3(\bm{x},\bm{z}), 
\label{mu_static} 
\end{equation} 
and in this case the potential, as given by
Eq.~(\ref{Phi_Poisson_sol}), becomes 
\begin{equation} 
\Phi(\bm{x}) = q G_3(\bm{x},\bm{z}). 
\label{Phi_static} 
\end{equation} 
The link between Eqs.~(\ref{Phi_gamma}) and (\ref{Phi_static}) can be
seen directly from Eq.~(\ref{G3_and_G4}). 

In the conformal formulation we have instead 
\begin{equation} 
\tilde{\mu}(\bm{x}) = q 
\frac{\tilde{\delta}_3(\bm{x},\bm{z})}{\Omega(\bm{z})},  
\end{equation} 
and Eq.~(\ref{Phi_Psol_conf}) produces 
\begin{equation} 
\Phi(\bm{x}) = q \frac{\tilde{G}_3(\bm{x},\bm{z})}
{\Omega(\bm{z})}.  
\label{Phi_static_conf} 
\end{equation} 
This result is compatible with Eq.~(\ref{Phi_static}) by virtue of
Eq.~(\ref{G3tilde_vs_G3}).  

\subsection{Electromagnetic field} 
\label{subsec:fg_em} 

A current density $j^\alpha$ creates an electromagnetic field 
$F_{\alpha\beta}$ that satisfies Maxwell's equations 
\begin{equation} 
F^{\alpha\beta}_{\ \ \ ;\beta} = 4\pi j^\alpha, \qquad 
F_{\alpha\beta;\gamma} + F_{\gamma\alpha;\beta} 
+ F_{\beta\gamma;\alpha} = 0. 
\label{Maxwell} 
\end{equation} 
The homogeneous equations are automatically satisfied when the
field tensor is expressed in terms of a 
vector potential $\Phi_{\alpha}$, 
\begin{equation} 
F_{\alpha\beta} = \nabla_\alpha \Phi_{\beta} 
- \nabla_\beta \Phi_{\alpha}.  
\end{equation} 
The inhomogeneous equations then take the form of a wave equation for
the vector potential, 
\begin{equation} 
\Box \Phi_{\alpha} - R_{\alpha}^{\ \beta} \Phi_\beta 
= -4\pi j_\alpha, 
\label{em_wave} 
\end{equation} 
provided that $\Phi_\alpha$ is required to satisfy the Lorenz gauge 
condition 
\begin{equation} 
\nabla_\alpha \Phi^\alpha = 0.
\label{Lorenz} 
\end{equation} 

In a static spacetime, and for a static distribution of charge, the
only relevant component of Maxwell's equations is 
$F^{t\beta}_{\ \ \ ;\beta} = 4\pi j^t$, and with 
$F_{ta} := -\partial_a \Phi_t$ this reduces to 
\begin{equation} 
\nabla^2 \Phi_t - A^a \partial_a \Phi_t = 4\pi \mu, \qquad 
\mu := N^2 j^t = -j_t. 
\label{Phit_Poisson} 
\end{equation}  
This equation also follows from evaluating Eq.~(\ref{em_wave}) in a
static spacetime. Comparing with Eq.~(\ref{Phi_Poisson}), we see that
$\Phi_t$ satisfies the same Poisson equation as a scalar potential
$\Phi$, except that the sign of $A^a$ is reversed. It is easy to see
that the gauge condition of Eq.~(\ref{Lorenz}) becomes 
$D_a \Phi^a + A_a \Phi^a = 0$ in a static spacetime, and that it has
no impact on $\Phi_t$.  

In a conformal formulation in which the spatial metric is expressed as
$h_{ab} = \Omega^2 \tilde{h}_{ab}$, Eq.~(\ref{Phit_Poisson}) becomes 
\begin{equation} 
\tilde{\nabla}^2 \Phi_t - \tilde{A}^a \partial_a \Phi_t 
= 4\pi \tilde{\mu}, 
\label{Phit_P_conformal} 
\end{equation}  
in which 
\begin{equation} 
\tilde{A}^a := \tilde{h}^{ab} \partial_b \ln\biggl( \frac{N}{\Omega}
\biggr) = \tilde{h}^{ab} \biggl( \frac{\partial_b N}{N} 
- \frac{\partial_b \Omega}{\Omega} \biggr) 
\label{em_Atilde} 
\end{equation} 
and 
\begin{equation} 
\tilde{\mu} = \Omega^2 \mu = N^2 \Omega^2 j^t. 
\label{em_mutilde} 
\end{equation}   

The four-dimensional Green's function associated with
Eq.~(\ref{em_wave}) is $G_\alpha^{\ \beta'}(x,x')$, which satisfies
the wave equation   
\begin{equation} 
\Box G_\alpha^{\ \beta'}(x,x') 
- R_\alpha^{\ \gamma} G_\gamma^{\ \beta'}(x,x') 
= -4\pi g_\alpha^{\ \beta'}(x,x') \delta_4(x,x'),  
\label{emG_wave} 
\end{equation} 
in which $g_\alpha^{\ \beta'}(x,x')$ is an operator of parallel
transport, taking a vector $A_{\beta'}$ at $x'$ and producing a
parallel-transported vector $A_\alpha$ at $x$. A solution to
Eq.~(\ref{em_wave}) is then given by  
\begin{equation} 
\Phi_\alpha(x) = \int G_\alpha^{\ \beta'}(x,x') j_{\beta'}(x')\, 
\sqrt{-g'} d^4x'. 
\label{em_wavesol} 
\end{equation} 

The three-dimensional Green's function associated with
Eq.~(\ref{Phit_Poisson})  is $G_3(\bm{x},\bm{x'})$, which satisfies the 
Poisson equation 
\begin{equation} 
\nabla^2 G_3(\bm{x},\bm{x'}) 
- A^a \partial_a G_3(\bm{x},\bm{x'}) = -4\pi
\delta_3(\bm{x},\bm{x'}). 
\label{emG3_Poisson} 
\end{equation} 
A solution to Eq.~(\ref{Phit_Poisson}) is
then given by 
\begin{equation} 
\Phi_t(\bm{x}) = -\int G_3(\bm{x},\bm{x'}) \mu(\bm{x'})\, 
\sqrt{h'} d^3x'.  
\label{Phit_Poisson_sol} 
\end{equation} 
Notice the minus sign on the right-hand side of
Eq.~(\ref{Phit_Poisson_sol}), which can be compared with its scalar
equivalent in Eq.~(\ref{Phi_Poisson_sol}). Notice also that while we
denote both the scalar and electromagnetic Green's functions by 
$G_3(\bm{x},\bm{x'})$, these functions are not equal to each other
because they satisfy distinct differential equations.  

The Green's function associated with Eq.~(\ref{Phit_P_conformal})  is 
$\tilde{G}_3(\bm{x},\bm{x'})$, which satisfies  
\begin{equation} 
\tilde{\nabla}^2 \tilde{G}_3(\bm{x},\bm{x'}) 
- \tilde{A}^a \partial_a \tilde{G}_3(\bm{x},\bm{x'}) 
= -4\pi \tilde{\delta}_3(\bm{x},\bm{x'}).  
\label{emG3_P_conf} 
\end{equation} 
A solution to Eq.~(\ref{Phit_P_conformal}) is then given by 
\begin{equation} 
\Phi_t(\bm{x}) = -\int \tilde{G}_3(\bm{x},\bm{x'})
\tilde{\mu}(\bm{x'})\, \sqrt{\tilde{h}'} d^3x',  
\label{Phit_Psol_conf} 
\end{equation} 
which features the same minus sign as in Eq.~(\ref{Phit_Poisson_sol}). 

The relation between $G_\alpha^{\ \beta'}(x,x')$ and
$G_3(\bm{x},\bm{x'})$ can be identified by performing
the time integration in Eq.~(\ref{em_wavesol}) and noticing that in 
a static situation, the integral involves $j_{t'}$ only. Comparing
with Eq.~(\ref{Phit_Poisson_sol}) produces   
\begin{equation} 
G_3(\bm{x},\bm{x'}) = \int G_t^{\ t'}(x,x')\, N(\bm{x'}) dt', 
\label{G3_and_Gem}
\end{equation} 
essentially the same relation as in the scalar case.  

The relation between $\tilde{G}_3$ and $G_3$ is found by making the 
substitutions $h^{1/2} = \Omega^3 \tilde{h}^{1/2}$ and $\mu =
\Omega^{-2} \tilde{\mu}$ within Eq.~(\ref{Phit_Poisson_sol}) and
comparing with Eq.~(\ref{Phit_Psol_conf}). The result is 
\begin{equation} 
\tilde{G}_3(\bm{x},\bm{x'}) = \Omega(\bm{x'}) G_3(\bm{x},\bm{x'}), 
\label{emG3tilde_vs_emG3} 
\end{equation}   
the same relation as in the scalar case. 

When the current density $j^\alpha$ describes a point charge $e$
moving on a world line $\gamma$ described by the parametric relations
$z(\tau)$, we have that  
\begin{equation} 
j^\alpha(x) =e \int_\gamma g^\alpha_{\ \mu}(x,z) u^\mu 
\delta_4( x, z)\, d\tau. 
\label{j_point} 
\end{equation} 
For a general world line the electric charge produces a potential
given by Eq.~(\ref{em_wavesol}), which evaluates to 
\begin{equation} 
\Phi_\alpha(x) = e \int_\gamma G_{\alpha\mu}(x,z)
u^\mu\, d\tau. 
\label{em_gamma} 
\end{equation}  
For a static charge at a fixed position $\bm{z}$, the integral of
Eq.~(\ref{j_point}) evaluates to $j^t(\bm{x}) = e N^{-1}(\bm{z})
\delta_3(\bm{x},\bm{z})$, so that   
\begin{equation} 
\mu(\bm{x}) = e N(\bm{z}) \delta_3(\bm{x},\bm{z}). 
\label{emmu_static}
\end{equation} 
In this case the potential, as given by
Eq.~(\ref{Phit_Poisson_sol}), becomes  
\begin{equation} 
\Phi_t(\bm{x}) = -e N(\bm{z}) G_3(\bm{x},\bm{z}).  
\label{Phit_static} 
\end{equation} 
Notice the extra minus sign and factor of $N$ when comparing this with
Eq.~(\ref{Phi_static}). In the conformal formulation we have instead  
\begin{equation} 
\tilde{\mu}(\bm{x}) = e \frac{N(\bm{z})}{\Omega(\bm{z})}\,   
\tilde{\delta}_3(\bm{x},\bm{z}),  
\end{equation} 
and Eq.~(\ref{Phit_Psol_conf}) produces 
\begin{equation} 
\Phi_t(\bm{x}) = -e \frac{N(\bm{z})}{\Omega(\bm{z})} 
\tilde{G}_3(\bm{x},\bm{z}). 
\label{Phit_static_conf} 
\end{equation} 
This result is compatible with Eq.~(\ref{Phit_static}) by virtue of
Eq.~(\ref{emG3tilde_vs_emG3}).  

\section{Hadamard's construction} 
\label{sec:Hadamard} 

\subsection{Scalar field} 
\label{subsec:H_scalar} 

We wish to find a representation for a Green's function
$G_3(\bm{x},\bm{x'})$ that satisfies 
\begin{equation} 
\nabla^2 G_3(\bm{x},\bm{x'}) 
+ A^a \partial_a G_3(\bm{x},\bm{x'}) = -4\pi \delta_3(\bm{x},\bm{x'}).  
\label{G3_P2} 
\end{equation} 
The Hadamard's construction developed here applies to this equation
--- a copy of Eq.~(\ref{G3_Poisson}) --- but it applies just as well
to the conformally-related formulation of Eq.~(\ref{G3_P_conf}).  

The Hadamard construction for the Green's function is
\cite{hadamard:23} 
\begin{equation} 
G^{\sf H}_3(\bm{x},\bm{x'}) =\frac{W(\bm{x},\bm{x'})}
{\sqrt{2\sigma(\bm{x},\bm{x'})}}, 
\label{G3_Hadamard} 
\end{equation} 
in which the two-point function $W(\bm{x},\bm{x'})$ satisfies the
differential equation 
\begin{equation} 
2 \sigma^a \partial_a W 
+ \bigl( \nabla^2 \sigma + A^a \sigma_a - 3 \bigr) W    
- (2\sigma) \bigl( \nabla^2 W + A^a \partial_a W \bigr) = 0 
\label{W_diffeq} 
\end{equation} 
together with the boundary condition 
\begin{equation} 
W(\bm{x'},\bm{x'}) = 1; 
\label{W_boundary} 
\end{equation} 
here $\sigma_a := \partial \sigma/\partial x^a$ and $\nabla^2 \sigma 
:= h^{ab} D_{a} \sigma_b$. The two-point function is known to be
smooth in the coincidence limit $\bm{x} \to \bm{x'}$, so that the
factor $(2\sigma)^{-1/2}$ is fully responsible for the singular
behavior of the Green's function at coincidence.   

To construct $W$ we express it as an expansion in powers of $2\sigma$,   
\begin{equation} 
W(\bm{x},\bm{x'}) = \sum_{n=0}^\infty W_n(\bm{x},\bm{x'}) 
\bigl[ 2\sigma(\bm{x},\bm{x'}) \bigr]^n, 
\label{W_expansion} 
\end{equation}
insert this within Eq.~(\ref{W_diffeq}), and collect powers of
$2\sigma$, making use of the identity $\sigma^a \sigma_a = 2\sigma$. 
Setting each coefficient to zero, we find that each $W_n$ must satisfy
the differential equation   
\begin{align} 
\nabla^2 W_{n-1} + A^a \partial_a W_{n-1} &= 
2(1-2n) \sigma^a \partial_a W_n 
\nonumber \\ & \hspace*{-60pt} \mbox{} 
+ (1-2n) \bigl( \nabla^2 \sigma + A^a \sigma_a \bigr) W_n 
\nonumber \\ & \hspace*{-60pt} \mbox{} 
- \bigl[ 3 + 4n(n-2) \bigr] W_n.
\label{Wn_diffeq} 
\end{align} 
The burden of enforcing Eq.~(\ref{W_boundary}) is then placed
solely upon $W_0$, which must satisfy 
\begin{equation} 
W_0(\bm{x'},\bm{x'}) = 1. 
\label{W0_boundary} 
\end{equation} 
Equation (\ref{Wn_diffeq}) is a recursion relation for each
$W_n$. With $W_{n-1}$ previously determined, $W_n$ is
obtained by selecting a base point $\bm{x'}$ and integrating
Eq.~(\ref{Wn_diffeq}) along each geodesic that emanates from
$\bm{x'}$. 

\subsection{Electromagnetic field} 
\label{subsec:H_em} 

We now wish to find the Hadamard representation for the
electromagnetic Green's function $G_3(\bm{x},\bm{x'})$, which
satisfies  
\begin{equation} 
\nabla^2 G_3(\bm{x},\bm{x'}) 
- A^a \partial_a G_3(\bm{x},\bm{x'}) = -4\pi \delta_3(\bm{x},\bm{x'}).  
\end{equation} 
The Hadamard construction applies to this equation --- a copy of 
Eq.~(\ref{emG3_Poisson}) --- but it applies just as well to the
conformally-related formulation of Eq.~(\ref{emG3_P_conf}).    

The construction is obtained directly from the scalar case by altering  
the sign of $A^a$ in all equations. The Hadamard representation for
the Green's function is   
\begin{equation} 
G^{\sf H}_3(\bm{x},\bm{x'}) =\frac{W(\bm{x},\bm{x'})}
{\sqrt{2\sigma(\bm{x},\bm{x'})}}, 
\label{emG3_Hadamard} 
\end{equation} 
in which the two-point function $W(\bm{x},\bm{x'})$ is smooth in the
coincidence limit $\bm{x} \to \bm{x'}$. It satisfies the differential
equation  
\begin{equation} 
2 \sigma^a \partial_a W 
+ \bigl( \nabla^2 \sigma - A^a \sigma_a - 3 \bigr) W    
- (2\sigma) \bigl( \nabla^2 W - A^a \partial_a W \bigr) = 0 
\label{emW_diffeq} 
\end{equation} 
together with the boundary condition 
\begin{equation} 
W(\bm{x'},\bm{x'}) = 1. 
\label{emW_boundary} 
\end{equation} 
As in the scalar case we express $W$ as an expansion in 
powers of $2\sigma$,  
\begin{equation} 
W(\bm{x},\bm{x'}) = \sum_{n=0}^\infty W_n(\bm{x},\bm{x'}) 
\bigl[ 2\sigma(\bm{x},\bm{x'}) \bigr]^n, 
\label{emW_expansion} 
\end{equation}
in which each coefficient $W_n$ must satisfy the
differential equation  
\begin{align} 
\nabla^2 W_{n-1} - A^a \partial_a W_{n-1} &= 
2(1-2n) \sigma^a \partial_a W_n 
\nonumber \\ & \hspace*{-60pt} \mbox{} 
+ (1-2n) \bigl( \nabla^2 \sigma - A^a \sigma_a \bigr) W_n 
\nonumber \\ & \hspace*{-60pt} \mbox{} 
- \bigl[ 3 + 4n(n-2) \bigr] W_n.
\label{emWn_diffeq} 
\end{align} 

\section{Local expansion of Hadamard's function} 
\label{sec:local} 

\subsection{Scalar field} 
\label{subsec:local_scalar} 

We wish to express the three-dimensional Green's function
$G^{\sf H}_3(\bm{x},\bm{x'})$ as a local expansion about the base
point $\bm{x'}$. We return to the Hadamard construction of
Eq.~(\ref{G3_Hadamard}) with the expansion of Eq.~(\ref{W_expansion}),
and now express $W_0$ and $W_1$ as the local expansions 
\begin{align} 
W_0 &= 1 + W^0_{a'} \sigma^{a'} 
+ \frac{1}{2} W^0_{a'b'} \sigma^{a'} \sigma^{b'}   
+ \frac{1}{6} W^0_{a'b'c'} \sigma^{a'} \sigma^{b'} \sigma^{c'}   
\nonumber \\ & \quad \mbox{} 
+ O(\epsilon^4) 
\label{W0_expansion} 
\end{align} 
and 
\begin{equation} 
W_1 = W^1 + W^1_{a'} \sigma^{a'} + O(\epsilon^2), 
\label{W1_expansion} 
\end{equation} 
in which $\sigma_{a'} := \partial \sigma/\partial x^{a'}$ and each
expansion coefficient is an ordinary tensor at $\bm{x'}$. We let
$\epsilon$ be a measure of the distance between $\bm{x}$ and
$\bm{x'}$, and the expansions of Eqs.~(\ref{W0_expansion}) and
(\ref{W1_expansion}) give rise to an expression for $W$ accurate
through order $\epsilon^3$. 

The expansion coefficients are determined by inserting
Eqs.~(\ref{W0_expansion}) and (\ref{W1_expansion}) within
Eq.~(\ref{Wn_diffeq}). We begin with $W_0$, which satisfies 
\begin{equation} 
2 \sigma{^a} \partial_a W_0 
+ \bigl( \nabla^2 \sigma + A^a \sigma_a - 3 \bigr) W_0 = 0. 
\label{W0_diffeq2} 
\end{equation} 
We rely on the standard expansions  
\begin{align} 
\sigma_{a b} &= h^{a'}_{\ a} h^{b'}_{\ b} \biggl[
h_{a'b'} - \frac{1}{3} R_{a'c'b'd'} \sigma^{c'} \sigma^{d'} 
\nonumber \\ & \quad \mbox{}  
+ \frac{1}{4} R_{a'c'b'd'|e'} \sigma^{c'} \sigma^{d'} \sigma^{e'} 
+ O(\epsilon^4) \biggr] 
\end{align} 
and 
\begin{equation} 
A_a = h^{a'}_{\ a} \biggl[ A_{a'} - A_{a'|c'} \sigma^{c'} 
+ \frac{1}{2} A_{a'|c'd'} \sigma^{c'} \sigma^{d'} 
+ O(\epsilon^3) \biggr],  
\end{equation} 
in which $\sigma_{ab} := D_a \sigma_b$ and 
$h^{a'}_{\ a}$ is the operator of parallel transport in the 
three-dimensional space. From the first equation we get 
\begin{equation} 
\nabla^2 \sigma = 3 - \frac{1}{3} R_{c'd'} \sigma^{c'} \sigma^{d'} 
+ \frac{1}{4} R_{c'd'|e'} \sigma^{c'} \sigma^{d'} \sigma^{e'} 
+ O(\epsilon^4), 
\end{equation} 
and the second equation gives rise to 
\begin{align} 
A_a \sigma^a &= -A_{c'} \sigma^{c'} 
+ A_{c'|d'} \sigma^{c'} \sigma^{d'} 
\nonumber \\ & \quad \mbox{}   
- \frac{1}{2} A_{c'|d'e'} \sigma^{c'} \sigma^{d'} \sigma^{e'}    
+ O(\epsilon^4)
\end{align} 
because $h^{a'}_{\ a} \sigma^{a} = -\sigma^{a'}$. Making use
of the identity $\sigma^{c'}_{\ a} \sigma^{a} = \sigma^{c'}$ we also
find that 
\begin{align} 
\sigma^{a} \partial_a W_0 &= W^0_{c'} \sigma^{c'} 
+ W^0_{c'd'} \sigma^{c'} \sigma^{c'}   
\nonumber \\ & \quad \mbox{}   
+ \frac{1}{2} W^0_{c'd'e'} \sigma^{c'} \sigma^{d'} \sigma^{e'}    
+ O(\epsilon^4). 
\end{align} 
Making the substitutions within Eq.~(\ref{W0_diffeq2}) and equating
each expansion coefficient to zero, we eventually arrive at
\begin{subequations} 
\label{W0_coefficients} 
\begin{align} 
W^0_{a'} &= \frac{1}{2} A_{a'}, \\ 
W^0_{a'b'} &= -\frac{1}{2} A_{a'|b'} + \frac{1}{4} A_{a'} A_{b'} 
+ \frac{1}{6} R_{a'b'}, \\  
W^0_{a'b'c'} &= \frac{1}{2} A_{(a'|b'c')} 
- \frac{3}{4} A_{(a'} A_{b'|c')} + \frac{1}{8} A_{a'} A_{b'} A_{c'} 
\nonumber \\ & \quad \mbox{}   
+ \frac{1}{4} A_{(a'} R_{b'c')} - \frac{1}{4} R_{(a'b'|c')}. 
\end{align} 
\end{subequations} 
It should be noted that since $A_{a'}$ is the gradient of a scalar
function, $A_{a'|b'} = A_{(a'|b')}$. 

We next turn to $W_1$, which satisfies the differential equation 
\begin{equation} 
2 \sigma^a \partial_a W_1 + \bigl( \nabla^2 \sigma 
+ A^a \sigma_a - 1 \bigr) W_1 = -\bigl( \nabla^2 W_0 
+ A^a \partial_a W_0 \bigr). 
\label{W1_diffeq} 
\end{equation} 
The left-hand side of the equation is computed with the same methods
as for the previous computation. For the right-hand side we make use
of the results $\sigma^{a'}_{\ b'} = \delta^{a'}_{\ b'} + O(\epsilon^2)$,
$\sigma^{a'}_{\ b} = -h^{a'}_{\ b} + O(\epsilon^2)$, and $\nabla^2
\sigma^{a'} = -\frac{2}{3} R^{a'}_{\ c'} \sigma^{c'} + O(\epsilon^2)$ to
obtain 
\begin{align} 
\nabla^2 W_0 &= h^{a'b'} W^0_{a'b'} + \biggl( 
-\frac{2}{3} W^0_{a'} R^{a'}_{\ c'} 
+ h^{a'b'} W^0_{a'b'c'} \biggr) \sigma^{c'} 
\nonumber \\ & \quad \mbox{}   
+ O(\epsilon^2) 
\end{align} 
and 
\begin{equation} 
A^a \partial_a W_0 = -A^{a'} W^0_{a'} - \bigl( 
A^{a'} W^0_{a'c'} - W^0_{a'} A^{a'}_{\ |c'} \bigr) \sigma^{c'} 
+ O(\epsilon^2).  
\end{equation}    
Making the substitutions within Eq.~(\ref{W1_diffeq}), equating
each expansion coefficient to zero, and simplifying the results with
Eq.~(\ref{W0_coefficients}), we eventually arrive at
\begin{subequations} 
\label{W1_coefficients} 
\begin{align} 
W^1 &= \frac{1}{4} A^{a'}_{\ |a'} + \frac{1}{8} A^{a'} A_{a'} 
- \frac{1}{12} R', \\ 
W^1_{a'} &= -\frac{1}{8} A^{c'}_{\ |c'a'} 
- \frac{1}{8} A^{c'} A_{c'|a'} 
+ \frac{1}{8} A^{c'}_{\ |c'} A_{a'} 
\nonumber \\ & \quad \mbox{}   
+ \frac{1}{16} A^{c'} A_{c'} A_{a'} 
- \frac{1}{24} R' A_{a'} 
+ \frac{1}{24} R_{|a'},  
\end{align} 
\end{subequations} 
in which $R'$ stands for the Ricci scalar evaluated at $\bm{x'}$. The
expression for $W^1_{a'}$ was simplified by invoking the contracted
Bianchi identity $R^{c'}_{\ a'|c'} = \frac{1}{2} R_{|a'}$ as well as
Ricci's identity to write $A_{a'\ \ c'}^{\ \ |c'} + A^{c'}_{\ |a'c'} 
+ A^{c'}_{\ |c'a'} = 3 A^{c'}_{\ |c'a'} + 2 R_{a'c'} A^{c'}$; recall
that $A_{a'|c'} = A_{c'|a'}$ because $A_{a'}$ is the gradient of a
scalar function.  

The local expansion of the Green's function is therefore 
\begin{align} 
G^{{\sf H}, {\rm local}}_3(\bm{x},\bm{x'}) 
&= \frac{1}{\sqrt{2\sigma}} \biggl\{  
1 + W^0_{a'} \sigma^{a'} 
+ \frac{1}{2} W^0_{a'b'} \sigma^{a'} \sigma^{b'}   
\nonumber \\ & \quad \mbox{}   
+ \frac{1}{6} W^0_{a'b'c'} \sigma^{a'} \sigma^{b'} \sigma^{c'}   
+ O(\epsilon^4) 
\nonumber \\ & \quad \mbox{} 
+ 2\sigma \Bigl[ W^1 + W^1_{a'} \sigma^{a'} 
+ O(\epsilon^2) \Bigr] 
\nonumber \\ & \quad \mbox{} 
+ O(\sigma^2) \biggr\}, 
\label{G3_local} 
\end{align} 
with the expansion coefficients listed in Eqs.~(\ref{W0_coefficients})
and (\ref{W1_coefficients}). 

\subsection{Electromagnetic field}
\label{subsec:local_em} 

We wish to express the three-dimensional Green's function
$G^{\sf H}_3(\bm{x},\bm{x'})$ as a local expansion about the base
point $\bm{x'}$. Once more we rely on the results from the scalar
case, which we directly import after implementing the substitution
$A^a \to -A^a$.   

The local expansion of the electromagnetic Green's function is 
\begin{align} 
G^{{\sf H}, {\rm local}}_3(\bm{x},\bm{x'}) 
&= \frac{1}{\sqrt{2\sigma}} \biggl\{  
1 + W^0_{a'} \sigma^{a'} 
+ \frac{1}{2} W^0_{a'b'} \sigma^{a'} \sigma^{b'}   
\nonumber \\ & \quad \mbox{} 
+ \frac{1}{6} W^0_{a'b'c'} \sigma^{a'} \sigma^{b'} \sigma^{c'}   
+ O(\epsilon^4) 
\nonumber \\ & \quad \mbox{} 
+ 2\sigma \Bigl[ W^1 + W^1_{a'} \sigma^{a'} 
+ O(\epsilon^2) \Bigr] 
\nonumber \\ & \quad \mbox{} 
+ O(\sigma^2) \biggr\}, 
\label{emG3_local} 
\end{align} 
with 
\begin{subequations} 
\label{emW0_coefficients} 
\begin{align} 
W^0_{a'} &= -\frac{1}{2} A_{a'}, \\ 
W^0_{a'b'} &= \frac{1}{2} A_{a'|b'} + \frac{1}{4} A_{a'} A_{b'} 
+ \frac{1}{6} R_{a'b'}, \\  
W^0_{a'b'c'} &= -\frac{1}{2} A_{(a'|b'c')} 
- \frac{3}{4} A_{(a'} A_{b'|c')} - \frac{1}{8} A_{a'} A_{b'} A_{c'} 
\nonumber \\ & \quad \mbox{} 
- \frac{1}{4} A_{(a'} R_{b'c')} - \frac{1}{4} R_{(a'b'|c')} 
\end{align} 
\end{subequations} 
and 
\begin{subequations} 
\label{emW1_coefficients} 
\begin{align} 
W^1 &= -\frac{1}{4} A^{a'}_{\ |a'} + \frac{1}{8} A^{a'} A_{a'} 
- \frac{1}{12} R', \\ 
W^1_{a'} &= \frac{1}{8} A^{c'}_{\ |c'a'} 
- \frac{1}{8} A^{c'} A_{c'|a'} 
+ \frac{1}{8} A^{c'}_{\ |c'} A_{a'} 
\nonumber \\ & \quad \mbox{} 
- \frac{1}{16} A^{c'} A_{c'} A_{a'} 
+ \frac{1}{24} R' A_{a'} 
+ \frac{1}{24} R_{|a'}.   
\end{align} 
\end{subequations} 

\section{Local expansion in conformal formulation} 
\label{sec:local_conformal} 

\subsection{Scalar field} 
\label{subsec:localconf_scalar} 

The local expansion of Eq.~(\ref{G3_local}) applies to the Hadamard 
representation of the Green's function $G_3(\bm{x},\bm{x'})$ defined by 
Eq.~(\ref{G3_Poisson}), but it applies just as well to the conformally
related Green's function $\tilde{G}_3(\bm{x},\bm{x'})$ defined by
Eq.~(\ref{G3_P_conf}); in this case one simply inserts the
conformally related quantities (such as $\tilde{A}^{a'}$,
$\tilde{\sigma}^{a'}$, and $\tilde{R}_{a'b'}$) in place of the
original quantities (such as $A^{a'}$, $\sigma^{a'}$, and
$R_{a'b'}$). As we shall now show, the expansions are then related by  
Eq.~(\ref{G3tilde_vs_G3}), 
\begin{equation} 
\tilde{G}^{{\sf H}, {\rm local}}_3(\bm{x},\bm{x'}) 
= \Omega(\bm{x'}) G^{{\sf H}, {\rm local}}_3(\bm{x},\bm{x'}), 
\label{G3tilde_vs_G3_copy} 
\end{equation} 
a conclusion that guarantees the consistency of the two approaches 
to the local expansion. Thus, a local expansion formulated in the
original space, and a local expansion formulated in the conformally
related space, will produce the same Green's function, apart from the
factor of $\Omega(\bm{x'})$ that appears in the relationship between
the Green's functions. 

This conclusion can be verified by straightforward computation, making
use of the well-known relations between conformally related
quantities. These include 
\begin{widetext} 
\begin{subequations} 
\label{conf_transf} 
\begin{align} 
\tilde{A}_a &= A_a + B_a, \\ 
\tilde{h}_{ab} &= \Omega^{-2} h_{ab}, \\  
\tilde{\Gamma}^a_{\, bc} &= \Gamma^a_{\, bc} + \delta^a_{\ b} B_c  
+ \delta^a_{\ c} B_b - h_{bc} B^a, \\ 
\tilde{R}^a_{\ bcd} &= R^a_{\ bcd} + \delta^a_{\ c} D_d B_b 
- \delta^a_{\ d} D_c B_b - h_{bc} D_d B^a + h_{bd} D_c B^a 
+ \delta^a_{\ c} B_b B_d - \delta^a_{\ d} B_b B_c 
\nonumber \\ & \quad \mbox{}
- \delta^a_{\ c} h_{bd} B_m B^m + \delta^a_{\ d} h_{bc} B_m B^m
- h_{bc} B^a B_d + h_{bd} B^a B_c, \\ 
\tilde{R}_{ab} &= R_{ab} + D_a B_b + h_{ab} D_m B^m + B_a B_b 
- h_{ab} B_m B^m, \\ 
\tilde{R} &= \Omega^2 \Bigl( R + 4 D_m B^m - 2 B_m B^m \Bigr), \\ 
\tilde{D}_c \tilde{R}_{ab} &= D_c R_{ab} + D_c D_a B_b 
+ h_{ab} D_c D_m B^m + 2\bigl( B_a D_b B_c + B_c D_a B_b 
+ B_b D_c B_a \bigr) 
\nonumber \\ & \quad \mbox{}
- \bigl( h_{ac} B^m D_m B_b 
+ h_{bc} B^m D_m B_a + 2 h_{ab} B^m D_m B_c \bigr) 
+ 2 h_{ab} B_c D_m B^m 
\nonumber \\ & \quad \mbox{}
+ 2 R_{ab} B_c + R_{ac} B_b + R_{bc} B_a 
- h_{ac} R_{bm} B^m - h_{bc} R_{am} B^m + 4 B_a B_b B_c 
\nonumber \\ & \quad \mbox{}
- \bigl( h_{a c} B_b + h_{b c} B_a + 2 h_{ab} B_c \bigr) B_m B^m, \\ 
\tilde{D}_a \tilde{R} &= \Omega^2 \Bigl( D_a R + 4 D_a D_m B^m 
- 4 B^m D_a B_m + 8 B_a D_m B^m + 2 R B_a - 4 B_a B_m B^m \Bigr), 
\end{align} 
\end{subequations} 
in which $B_a := \partial_a \ln \Omega$, and where all indices on the
right-hand side are raised with $h^{ab}$. They include also 
\begin{equation} 
\tilde{\sigma}^{a'} = \sigma^{a'} 
+ \frac{1}{2} S^{a'}_{\ b'c'} \sigma^{a'} \sigma^{b'} \sigma^{c'} 
+ \frac{1}{6} S^{a'}_{\ b'c'd'} \sigma^{a'} \sigma^{b'} \sigma^{c'}
\sigma^{d'} + \frac{1}{24} S^{a'}_{\ b'c'd'e'} \sigma^{a'} \sigma^{b'}
\sigma^{c'} \sigma^{d'} \sigma^{e'} + O(\epsilon^6), 
\end{equation} 
an expansion of $\tilde{\sigma}^{a'} := \tilde{h}^{a'b'}
\tilde{D}_{b'} \tilde{\sigma}$ in powers of $\sigma^{a'} := h^{a'b'}
D_{b'} \sigma$, in which $\tilde{\sigma}$ is half the geodetic
separation in the conformally related space. The expansion 
coefficients are given by \cite{tadaki:89} 
\begin{subequations} 
\begin{align} 
S^{a'}_{\ b'c'} &= 2 \delta^{a'}_{\ b'} B_{c'} - h_{b'c'} B^{a'}, \\ 
S^{a'}_{\ b'c'd'} &= -2 \delta^{a'}_{\ b'} D_{c'} B_{d'} 
+ h_{b'c'} D_{d'} B^{a'} + 4 \delta^{a'}_{\ b'} B_{c'} B_{d'} 
- \delta^{a'}_{\ b'} h_{c'd'} B_{m'} B^{m'} 
- 2 g_{b'c'} B^{a'} B_{d'}, \\ 
S^{a'}_{\ b'c'd'e'} &= 2 \delta^{a'}_{\ b'} D_{c'} D_{d'} B_{e'} 
- h_{b'c'} D_{d'} D_{e'} B^{a'} 
- 12 \delta^{a'}_{\ b'} B_{c'} D_{d'} B_{e'} 
+ 4 h_{b'c'} B_{d'} D_{e'} B^{a'} 
+ 4 \delta^{a'}_{\ b'} h_{c'd'} B^{m'} D_{e'} B_{m'} 
\nonumber \\ & \quad \mbox{}
- h_{b'c'} h_{d'e'} B^{m'} D_{m'} B^{a'} 
+ 2 h_{b'c'} B^{a'} D_{d'} B_{e'} 
+ 8 \delta^{a'}_{\ b'} B_{c'} B_{d'} B_{e'} 
- 4 \delta^{a'}_{\ b'} h_{c'd'} B_{e'} B_{m'} B^{m'} 
\nonumber \\ & \quad \mbox{}
- 4 h_{b'c'} B_{d'} B_{e'} B^{a'} 
+ h_{b'c'} h_{d'e'} B_{m'} B^{m'} B^{a'} 
- h_{b'c'} R^{a'}_{\ d'm'e'} B^{m'}, 
\end{align} 
\end{subequations} 
in which the lower indices $b'c'$, or $b'c'd'$, or $b'c'd'e'$ are
understood to be fully symmetrized on the right-hand side of the
equations. These relations imply 
\begin{equation} 
\tilde{\sigma}(\bm{x},\bm{x'}) = 
\Omega^{-2}(\bm{x'}) \sigma(\bm{x},\bm{x'}) \biggl[ 
1 + P_{a'} \sigma^{a'} + \frac{1}{2} P_{a'b'} \sigma^{a'} \sigma^{b'} 
+ \frac{1}{6} P_{a'b'c'} \sigma^{a'} \sigma^{b'} \sigma^{c'} 
+ O(\epsilon^4) \biggr], 
\label{sigmatilde_vs_sigma} 
\end{equation}  
with 
\begin{subequations} 
\begin{align} 
P_{a'} &= B_{a'}, \\ 
P_{a'b'} &= -\frac{2}{3} D_{a'} B_{b'} + \frac{4}{3} B_{a'} B_{b'} 
- \frac{1}{6} h_{a'b'} B_{m'} B^{m'}, \\ 
P_{a'b'c'} &= \frac{1}{2} D_{a'} D_{b'} B_{c'} 
- 3 B_{a'} D_{b'} B_{c'} + \frac{1}{2} h_{a'b'} B^{m'} D_{c'} B_{m'} 
+ 2 B_{a'} B_{b'} B_{c'} - \frac{1}{2} h_{a'b'} B_{c'} B_{m'} B^{m'}, 
\end{align} 
\end{subequations} 
with the same understanding regarding the $a'b'$ or $a'b'c'$ indices
on the right-hand side.   
\end{widetext} 

To verify that Eq.~(\ref{G3tilde_vs_G3_copy}) holds, we begin with the
conformal formulation of Eq.~(\ref{G3_local}), in which we make the
substitutions listed above. Simplifying, and keeping all expansions
accurate through order $\epsilon^3$, reveals that indeed, the end
result is Eq.~(\ref{G3_local}) formulated in the original space,
except for the overall factor of $\Omega(\bm{x'})$ that occurs in 
Eq.~(\ref{G3tilde_vs_G3_copy}). Consistency of the local expansions is
therefore assured. 

It is natural to ask whether the validity of
Eq.~(\ref{G3tilde_vs_G3_copy}) could be established as an exact
relation, instead of as an approximate local expansion pursued through
order $\epsilon^3$. Defining a $\tilde{W}(\bm{x},\bm{x'})$ by the
relation 
\begin{equation} 
\tilde{W}(\bm{x},\bm{x'}) := \Omega(\bm{x'}) 
\sqrt{ \frac{\tilde{\sigma}(\bm{x},\bm{x'})}{\sigma(\bm{x},\bm{x'})} }
W(\bm{x},\bm{x'}),  
\end{equation} 
the proof would amount to a demonstration that this $\tilde{W}$ is
suitable to be implicated in a Hadamard construction of the conformal
Green's function via $\tilde{G}^{\sf H}_3 
= \tilde{W}/\sqrt{2\tilde{\sigma}}$.    

The proof would involve three essential steps. First, the function
$\tilde{W}$, as defined here, must be shown to satisfy the same
differential equation as Eq.~(\ref{W_diffeq}) expressed in its 
conformal formulation; this property follows directly from the fact
that $\tilde{W} = \sqrt{2\tilde{\sigma}}\, \tilde{G}_3$, in which the
two-point function $\tilde{G}_3 := \Omega(\bm{x'}) G_3^{\sf H}$ 
is known to satisfy Eq.~(\ref{G3_P_conf}), the conformal
formulation of Green's equation. (The issue at stake is whether this 
$\tilde{G}_3$, which is defined as $\tilde{W}/\sqrt{2\tilde{\sigma}}$,
is a proper Hadamard representation of the conformal Green's
function.)  Second, $\tilde{W}$ must be shown to satisfy the boundary
condition of Eq.~(\ref{W_boundary}); this property follows immediately
from the coincidence limit of Eq.~(\ref{sigmatilde_vs_sigma}) and the
fact that $W$ itself satisfies the boundary condition. Third,
$\tilde{W}$ must be shown to be {\it smooth} at 
$\bm{x} = \bm{x'}$, by which we mean that the function must be
$C^\infty$ when viewed as a function of $\bm{x}$ with $\bm{x'}$
fixed; this property ensures that $\tilde{W}$ admits an expansion in
powers of $\tilde{\sigma}$ as displayed in Eq.~(\ref{W_expansion}),
which is known to be convergent and unique. The expansion being
unique, smoothness ensures that the Hadamard construction   
\begin{equation} 
\tilde{G}^{\sf H}_3(\bm{x},\bm{x'}) = \frac{ \tilde{W}(\bm{x},\bm{x'}) } 
{ \sqrt{2 \tilde{\sigma}(\bm{x},\bm{x'})} } 
\end{equation} 
gives rise to a Green's function that satisfies
Eq.~(\ref{G3tilde_vs_G3_copy}) exactly. 

Evidence that $\tilde{W}$ is smooth through order $\epsilon^3$ was
presented in the context of the local expansion. Because $W$ is known
to be smooth, smoothness of $\tilde{W}$ to all orders relies on the
smoothness of $\tilde{\sigma}/\sigma$, which can only be assured if
the series expansion of Eq.~(\ref{sigmatilde_vs_sigma}) can be proved
to converge. In the absence of such a proof, we shall have to give the
exact version of Eq.~(\ref{G3tilde_vs_G3_copy}) the status of a
plausible, but unproved, conjecture.   

\subsection{Electromagnetic field} 
\label{subsec:localconf_em} 

In Sec.~\ref{subsec:localconf_scalar}  we were able to establish
that in the case of a scalar field, a local expansion of the Hadamard
Green's function formulated in the original space, and a local
expansion formulated in the conformally related space, produce the
same Green's function, apart from the factor of $\Omega(\bm{x'})$ that
appears in Eq.~(\ref{emG3tilde_vs_emG3}). In addition, we formulated a
conjecture to the effect that the two Hadamard forms may be related by 
\begin{equation} 
\tilde{G}^{\sf H}_3(\bm{x},\bm{x'}) 
= \Omega(\bm{x'}) G_3^{\sf H}(\bm{x},\bm{x'})     
\end{equation} 
as a matter of exact identity. The methods of
Sec.~\ref{subsec:localconf_scalar} allow us to make the same
statements regarding the electromagnetic Green's function. The
required computations are almost identical, and all the relevant
equations can be obtained from the scalar case by making the
substitution $A^a \to -A^a$. 

\section{Detweiler-Whiting construction} 
\label{sec:DW} 

\subsection{Scalar field} 
\label{subsec:DW_scalar} 

In this section we construct the three-dimensional version of the
Detweiler-Whiting singular Green's function for a static scalar field
in a static spacetime. By virtue of Eq.~(\ref{G3_and_G4}), this can be
related to the four-dimensional version of Eq.~(\ref{DW_def}) by 
\begin{equation} 
G^{\sf S}_3(\bm{x},\bm{z}) := \int_\gamma G^{\sf S}_4(x,z)\, d\tau  
\label{G3S_def1}
\end{equation} 
in which $\tau$ is proper time for an observer at rest at the spatial
position $\bm{z}$. The integral can be evaluated with the techniques
described in Sec.~17.2 of Ref.~\cite{poisson-pound-vega:11}, and we
have that   
\begin{align} 
G^{\sf S}_3(\bm{x},\bm{\bar{x}}) &= 
\frac{1}{2r} U(x,x') + \frac{1}{2r_{\rm adv}} U(x,x'') 
\nonumber \\ & \quad \mbox{} 
- \frac{1}{2} \int_u^v V(x,z)\, d\tau, 
\label{G3S_def2}
\end{align} 
in which $x' := z(u)$ is the retarded point on the (static) world
line, $x'' := z(v)$ is the advanced point, 
$r := \sigma_{\alpha'} u^{\alpha'}$ is the retarded distance, 
$r_{\rm adv} := -\sigma_{\alpha''} u^{\alpha''}$ is the advanced
distance, and $U(x,z)$ and $V(x,z)$ are the two-point functions that
appear in the construction of the four-dimensional Green's function.   

To calculate $G^{\sf S}_3$ we follow the methods of Haas and Poisson
(HP) \cite{haas-poisson:06}, wherein the retarded and advanced points
are related to a middle point $\bar{x}$ on the world line. But while
$\bar{x}$ was chosen arbitrarily in HP, here we specifically choose
$\bar{x}$ to be simultaneous with $x$, so that $\bar{x}$ and $x$ have
the same time coordinate. This condition implies that 
$\bar{r} := \sigma_{\bar{\alpha}}(x,\bar{x}) u^{\bar{\alpha}}= 0$.

Following HP we define the world-line functions 
\begin{subequations} 
\begin{align} 
\sigma(\tau) &:= \sigma\bigl(x,z(\tau)\bigr), \\ 
U(\tau) &:= U\bigl(x,z(\tau)\bigr), \\ 
V(\tau) &:= V\bigl(x,z(\tau)\bigr),
\end{align} 
\end{subequations} 
in which $x$ is kept fixed. These functions will all be expressed
as Taylor expansions about $\tau = \bar{\tau}$, with $\bar{\tau}$
defined by $\bar{x} := z(\bar{\tau})$. We also define  
\begin{equation} 
s^2 := g^{\bar{\alpha}\bar{\beta}} 
\sigma_{\bar{\alpha}} \sigma_{\bar{\beta}} 
= 2 \sigma(x,\bar{x}),  
\end{equation} 
the squared geodesic distance between $x$ and $\bar{x}$. Notice that
$r =\dot{\sigma}(u)$ and $r_{\rm adv} = -\dot{\sigma}(v)$, in which an
overdot indicates differentiation with respect to $\tau$. We define  
\begin{equation} 
\Delta_- := u - \bar{\tau}, \qquad
\Delta_+ := v - \bar{\tau}, 
\end{equation} 
with $\Delta_- < 0$ and $\Delta_+ > 0$; these parameters are
collectively denoted $\Delta$. 

The $\Delta$ parameters are determined by writing $\sigma(u) = 0$ or
$\sigma(v) = 0$ as a Taylor expansion about $\bar{\tau}$: 
\begin{align} 
0 &= \sigma + \dot{\sigma} \Delta + \frac{1}{2} \ddot{\sigma} \Delta^2 
+ \frac{1}{6} \dddot{\sigma} \Delta^3 
+ \frac{1}{24} \sigma^{(4)} \Delta^4  
\nonumber \\ & \quad \mbox{}
+ \frac{1}{120} \sigma^{(5)} \Delta^5 + O(\epsilon^6),  
\label{Delta1} 
\end{align} 
in which $\sigma$ and its derivatives are evaluated at $\tau =
\bar{\tau}$. This equation is then solved for $\Delta$. The
derivatives of $\sigma(\tau)$ are given by
\begin{widetext}  
\begin{subequations} 
\begin{align} 
\sigma &= \frac{1}{2} s^2, \\ 
\dot{\sigma} &= \sigma_{\bar{\alpha}} u^{\bar{\alpha}} = 0, \\ 
\ddot{\sigma} &= \sigma_{\bar{\alpha}\bar{\beta}} u^{\bar{\alpha}}
u^{\bar{\beta}} + \sigma_{\bar{\alpha}} a^{\bar{\alpha}}
\nonumber \\ 
&= -1 - \frac{1}{3} R_{u\sigma u\sigma} 
+ \frac{1}{12} R_{u\sigma u\sigma;\sigma} 
+\sigma_{\bar{\alpha}} a^{\bar{\alpha}}+ O(\epsilon^4), \\ 
\dddot{\sigma} &= \sigma_{\bar{\alpha}\bar{\beta}\bar{\gamma}}
u^{\bar{\alpha}} u^{\bar{\beta}} u^{\bar{\gamma}} 
+ 3\sigma_{\bar{\alpha}\bar{\beta}} u^{\bar{\alpha}} a^{\bar{\beta}} 
+  \sigma_{\bar{\alpha}} \dot{a}^{\bar{\alpha}} 
\nonumber \\
&= -\frac{1}{4} R_{u\sigma u\sigma;u} - R_{u\sigma a\sigma} 
+ \sigma_{\bar{\alpha}} \dot{a}^{\bar{\alpha}} + O(\epsilon^3), \\ 
\sigma^{(4)} &=
\sigma_{\bar{\alpha}\bar{\beta}\bar{\gamma}\bar{\delta}} 
u^{\bar{\alpha}} u^{\bar{\beta}} u^{\bar{\gamma}} u^{\bar{\delta}} 
+ \sigma_{\bar{\alpha}\bar{\beta}\bar{\gamma}} 
\bigl( 5u^{\bar{\alpha}} a^{\bar{\beta}} u^{\bar{\gamma}} 
+ u^{\bar{\alpha}} u^{\bar{\beta}} a^{\bar{\gamma}} \bigr) 
+ \sigma_{\bar{\alpha}\bar{\beta}} 
\bigl( 3a^{\bar{\alpha}} a^{\bar{\beta}} 
+ 4 u^{\bar{\alpha}} \dot{a}^{\bar{\beta}} \bigr) 
+ \sigma_{\bar{\alpha}} \ddot{a}^{\bar{\alpha}}
\nonumber \\
&= R_{uau\sigma} - a^2 + \sigma_{\bar{\alpha}} \ddot{a}^{\bar{\alpha}} 
+ O(\epsilon^2), \\ 
\sigma^{(5)} &=
\sigma_{\bar{\alpha}\bar{\beta}\bar{\gamma}\bar{\delta}\bar{\epsilon}} 
u^{\bar{\alpha}} u^{\bar{\beta}} u^{\bar{\gamma}} u^{\bar{\delta}}
u^{\bar{\epsilon}} 
+ \sigma_{\bar{\alpha}\bar{\beta}\bar{\gamma}\bar{\delta}}  
\bigl( 
a^{\bar{\alpha}} u^{\bar{\beta}} u^{\bar{\gamma}} u^{\bar{\delta}} 
+ 6 u^{\bar{\alpha}} a^{\bar{\beta}} u^{\bar{\gamma}} u^{\bar{\delta}} 
+ 2 u^{\bar{\alpha}} u^{\bar{\beta}} a^{\bar{\gamma}} u^{\bar{\delta}} 
+ u^{\bar{\alpha}} u^{\bar{\beta}} u^{\bar{\gamma}} a^{\bar{\delta}} 
\bigr) 
\nonumber \\ & \quad \mbox{} 
+ \sigma_{\bar{\alpha}\bar{\beta}\bar{\gamma}} 
\bigl( 8 a^{\bar{\alpha}} a^{\bar{\beta}} u^{\bar{\gamma}}
+ 6 u^{\bar{\alpha}} a^{\bar{\beta}} a^{\bar{\gamma}}
+ a^{\bar{\alpha}} u^{\bar{\beta}} a^{\bar{\gamma}}
+ 9 u^{\bar{\alpha}} \dot{a}^{\bar{\beta}} u^{\bar{\gamma}}
+ u^{\bar{\alpha}} u^{\bar{\beta}} \dot{a}^{\bar{\gamma}} \bigr) 
+ \sigma_{\bar{\alpha}\bar{\beta}} 
\bigl( 10 a^{\bar{\alpha}} \dot{a}^{\bar{\beta}} 
+ 5 u^{\bar{\alpha}} \ddot{a}^{\bar{\beta}} \bigr) 
+ \sigma_{\bar{\alpha}} \dddot{a}^{\bar{\alpha}} 
\nonumber \\
&= -5 a_{\bar{\alpha}} \dot{a}^{\bar{\alpha}} + O(\epsilon). 
\end{align} 
\end{subequations} 
\end{widetext} 
These results rely on the standard expansion 
\begin{align} 
\sigma_{\bar{\alpha}\bar{\beta}} (x,\bar{x}) &= 
g_{\bar{\alpha}\bar{\beta}} 
- \frac{1}{3} R_{\bar{\alpha}\bar{\mu}\bar{\beta}\bar{\nu}} 
\sigma^{\bar{\mu}} \sigma^{\bar{\nu}} 
\nonumber \\ & \quad \mbox{}
+ \frac{1}{12}
R_{\bar{\alpha}\bar{\mu}\bar{\beta}\bar{\nu};\bar{\lambda}}  
\sigma^{\bar{\mu}} \sigma^{\bar{\nu}} \sigma^{\bar{\lambda}} 
+ O(\epsilon^4), 
\end{align} 
which can be differentiated with respect to $\bar{x}^\alpha$ to
produce expansions for 
$\sigma_{\bar{\alpha}\bar{\beta}\bar{\gamma}}$ and so on. We use the
HP notation for the components of the Riemann tensor; for example
$R_{u\sigma u\sigma} :=R_{\bar{\alpha}\bar{\mu}\bar{\beta}\bar{\nu}}  
u^{\bar{\alpha}} \sigma^{\bar{\mu}} u^{\bar{\beta}}
\sigma^{\bar{\nu}}$, $R_{u\sigma a\sigma} := 
R_{\bar{\alpha}\bar{\mu}\bar{\beta}\bar{\nu}}  
u^{\bar{\alpha}} \sigma^{\bar{\mu}} a^{\bar{\beta}}
\sigma^{\bar{\nu}}$, and $R_{u\sigma u\sigma;u} := 
R_{\bar{\alpha}\bar{\mu}\bar{\beta}\bar{\nu};\bar{\lambda}}   
u^{\bar{\alpha}} \sigma^{\bar{\mu}} u^{\bar{\beta}}
\sigma^{\bar{\nu}} u^{\bar{\lambda}}$. We have defined 
\begin{equation} 
a^\mu := \frac{D u^\mu}{d\tau}, \qquad 
\dot{a}^\mu := \frac{D a^\mu}{d\tau}, \qquad 
\ddot{a}^\mu := \frac{D \dot{a}^\mu}{d\tau},
\end{equation} 
and so on, and used the identities $u_\mu a^\mu = 0$, 
$u_\mu \dot{a}^\mu = -a^2 := a_\mu a^\mu$, and 
$u_\mu \ddot{a}^\mu = -3a_\mu \dot{a}^\mu$.  

Substitution of these expansions within Eq.~(\ref{Delta1}) and solving
for $\Delta$ returns an expansion of the form 
\begin{equation} 
\Delta = \Delta_1 \epsilon + \Delta_2 \epsilon^2
+ \Delta_3 \epsilon^3 + \Delta_4 \epsilon^4 + O(\epsilon^5). 
\label{Delta2} 
\end{equation} 
The explicit expressions for $\Delta_1$, $\Delta_2$, $\Delta_3$, and
$\Delta_4$ are too large to be displayed here, but we may mention that 
$\Delta^+_1 = s$ and $\Delta^-_1 = -s$.   

With $\Delta$ determined, $r$ and $r_{\rm adv}$ can be calculated as
Taylor expansions. Since $r = \dot{\sigma}(u)$ and $r_{\rm adv} =
-\dot{\sigma}(v)$, we have that   
\begin{subequations}
\label{r_Delta} 
\begin{align}  
r &= \ddot{\sigma} \Delta_- 
+ \frac{1}{2} \dddot{\sigma} \Delta_-^2  
+ \frac{1}{6} {\sigma}^{(4)} \Delta_-^3 
+ \frac{1}{24} {\sigma}^{(5)} \Delta_-^4 
\nonumber \\ & \quad \mbox{}
+ O(\epsilon^5), \\
r_{\rm adv} &= -\ddot{\sigma} \Delta_+ 
- \frac{1}{2} \dddot{\sigma} \Delta_+^2  
- \frac{1}{6} {\sigma}^{(4)} \Delta_+^3 
- \frac{1}{24} {\sigma}^{(5)} \Delta_+^4 
\nonumber \\ & \quad \mbox{}
+ O(\epsilon^5).  
\end{align}
\end{subequations}  
At leading order $r = s + O(\epsilon^2)$ and 
$r_{\rm adv} = s + O(\epsilon^2)$, but the complete expansions for 
$r^{-1}$ and $r^{-1}_{\rm adv}$ are too large to be displayed here. 

Expressions for $U(x,x')$ and $U(x,x'')$ are obtained in a similar
way. We write  
\begin{subequations}
\begin{align} 
U(x,x') &= U + \dot{U} \Delta_- + \frac{1}{2} \ddot{U} \Delta_-^2 
+ \frac{1}{6} \dddot{U} \Delta_-^3 
\nonumber \\ & \quad \mbox{}
+ O(\epsilon^4), \\
U(x,x'') &= U + \dot{U} \Delta_+ + \frac{1}{2} \ddot{U} \Delta_+^2 
+ \frac{1}{6} \dddot{U} \Delta_+^3 
\nonumber \\ & \quad \mbox{}
+ O(\epsilon^4), 
\end{align} 
\end{subequations} 
in which $U(\tau)$ and its derivatives are evaluated at $\tau =
\bar{\tau}$. These quantities are given by 
\begin{subequations} 
\begin{align} 
U &= 1 + \frac{1}{12} R_{\sigma\sigma} 
- \frac{1}{24} R_{\sigma\sigma;\sigma} +O(\epsilon^4), \\
\dot{U} &= U_{;\bar{\alpha}} u^{\bar{\alpha}}
\label{Uexp} 
\nonumber \\ 
&= \frac{1}{6} R_{u\sigma} + \frac{1}{24} R_{\sigma\sigma;u} 
- \frac{1}{12} R_{u\sigma;\sigma} +O(\epsilon^3), \\ 
\ddot{U} &= U_{;\bar{\alpha}\bar{\beta}} u^{\bar{\alpha}}
u^{\bar{\beta}} + U_{;\bar{\alpha}} a^{\bar{\alpha}}
\nonumber \\ 
&= \frac{1}{6} R_{uu} + \frac{1}{6} R_{a\sigma} 
+ \frac{1}{6} R_{u\sigma;u} - \frac{1}{12} R_{uu;\sigma}  
\nonumber \\ & \quad \mbox{}
+ O(\epsilon^2), \\ 
\dddot{U} &= U_{;\bar{\alpha}\bar{\beta}\bar{\gamma}} 
u^{\bar{\alpha}} u^{\bar{\beta}} u^{\bar{\gamma}} 
+ 3 U_{;\bar{\alpha}\bar{\beta}} a^{\bar{\alpha}}
u^{\bar{\beta}} + U_{;\bar{\alpha}} \dot{a}^{\bar{\alpha}}
\nonumber \\ 
&= \frac{1}{2} R_{au} + \frac{1}{4} R_{uu;u} 
+ O(\epsilon). 
\end{align} 
\end{subequations} 

To evaluate the tail integral we expand $V(\tau)$ as
$V(\bar{\tau}) + (\tau-\bar{\tau})\dot{V}(\bar{\tau}) 
+ O(\epsilon^2)$ and integrate with respect to $\tau$ between $u = 
\bar{\tau} + \Delta_-$ and $v = \bar{\tau} + \Delta_+$. The result is  
\begin{align} 
\int_u^v V(x,z)\, d\tau  &= V(\Delta_+ - \Delta_-) 
+ \frac{1}{2} \dot{V}(\Delta^2_+ - \Delta^2_-) 
\nonumber \\ & \quad \mbox{}
+ O(\epsilon^3), 
\end{align} 
in which $V := V(x,\bar{x})$ and 
$\dot{V} := V_{;\bar{\alpha}} u^{\bar{\alpha}}$. These are given by
the expansions  
\begin{equation} 
V = \frac{1}{12} \bar{R} 
- \frac{1}{24} R_{;\bar{\alpha}} \sigma^{\bar{\alpha}} 
+ O(\epsilon^2),
\label{Vexp} 
\end{equation}
and 
\begin{equation} 
\dot{V} = \frac{1}{24} R_{;\bar{\alpha}} u^{\bar{\alpha}} 
+ O(\epsilon). 
\end{equation} 
To obtain Eq.~(\ref{Vexp}) we rely on standard expansion
techniques. The two-point function is required to satisfy the wave
equation  $\Box V = 0$ as well as the light-cone equation
\begin{equation} 
V_{;\alpha} \sigma^\alpha 
+ \frac{1}{2} (\sigma^\alpha_{\ \alpha} - 2) V 
= \frac{1}{2} \Box U,  
\end{equation} 
which is evaluated at $\sigma(x,\bar{x}) = 0$. The solution is 
expressed as an expansion 
\begin{equation} 
V(x,\bar{x})  = \sum_{n=0}^\infty V_n(x,\bar{x}) \sigma^n, 
\end{equation} 
and the wave equation gives rise to a sequence of equations which
determine $V_n$ from $V_{n-1}$; the light-cone equation determines 
$V_0$. Because $\sigma = O(\epsilon^2)$, $V = V_0$ to order
$\epsilon$, and this can be obtained by inserting the expansion   
\begin{equation} 
V = V^0 + V^0_{\bar{\alpha}} \sigma^{\bar{\alpha}} + O(\epsilon^2) 
\end{equation} 
within the light-cone equation. We use the fact that  
$\sigma^\alpha_{\ \alpha} = 4 + O(\epsilon^2)$, and to compute  
$\Box U$ we start with Eq.~(\ref{Uexp}) and rely on the expansions  
\begin{equation} 
\sigma^{\bar{\alpha}}_{\ \alpha} = -g^{\bar{\alpha}}_{\ \alpha} 
+ O(\epsilon^2), \qquad 
g^{\bar{\alpha}}_{\ \alpha;\beta} = O(\epsilon); 
\end{equation} 
we eventually arrive at 
\begin{equation} 
\Box U = \frac{1}{6} \bar{R} 
- \frac{1}{6} R_{;\bar{\alpha}} \sigma^{\bar{\alpha}} 
+ O(\epsilon^2). 
\end{equation} 
The end result of the computation is Eq.~(\ref{Vexp}).  

Putting all the ingredients together, we eventually arrive at the
following expansion for $G^{\sf S}_3$: 
\begin{align} 
G^{\sf S}_3(\bm{x},\bm{\bar{x}}) &= \frac{1}{s} \biggl\{  
1 + \psi^0_{\bar{\alpha}} \sigma^{\bar{\alpha}}
+ \frac{1}{2} \psi^0_{\bar{\alpha}\bar{\beta}}
   \sigma^{\bar{\alpha}} \sigma^{\bar{\beta}}
\nonumber \\ & \quad \mbox{}
+ \frac{1}{6} \psi^0_{\bar{\alpha}\bar{\beta}\bar{\gamma}} 
   \sigma^{\bar{\alpha}} \sigma^{\bar{\beta}} \sigma^{\bar{\gamma}}  
+ O(\epsilon^4) 
\nonumber \\ & \quad \mbox{} 
+ s^2 \Bigl[ \psi^1 + \psi^1_{\bar{\alpha}} \sigma^{\bar{\alpha}} 
+ O(\epsilon^2) \Bigr] \biggr\}, 
\label{G3S_local1} 
\end{align} 
with 
\begin{subequations} 
\begin{align} 
\psi^0_{\bar{\alpha}} &= \frac{1}{2} a_{\bar{\alpha}}, \\ 
\psi^0_{\bar{\alpha}\bar{\beta}} &= 
\frac{3}{4} a_{\bar{\alpha}} a_{\bar{\beta}} 
+ \frac{1}{6} R_{\bar{\alpha}\bar{\beta}} 
- \frac{1}{3}u^{\bar{\mu}} u^{\bar{\nu}}
   R_{\bar{\mu}\bar{\alpha}\bar{\nu}\bar{\beta}}, \\
\psi^0_{\bar{\alpha}\bar{\beta}\bar{\gamma}} &= 
\frac{15}{8} a_{\bar{\alpha}} a_{\bar{\beta}} a_{\bar{\gamma}} 
- \frac{3}{2} a_{\bar{\alpha}} u^{\bar{\mu}} u^{\bar{\nu}}
   R_{\bar{\mu}\bar{\beta}\bar{\nu}\bar{\gamma}}
+ \frac{1}{4} a_{\bar{\alpha}} R_{\bar{\beta}\bar{\gamma}} 
\nonumber \\ & \quad \mbox{}
+ \frac{1}{4} u^{\bar{\mu}} u^{\bar{\nu}}
   R_{\bar{\mu}\bar{\alpha}\bar{\nu}\bar{\beta};\bar{\gamma}} 
- \frac{1}{4} R_{\bar{\alpha}\bar{\beta};\bar{\gamma}} 
\end{align} 
\end{subequations} 
and 
\begin{subequations} 
\begin{align} 
\psi^1 &= -\frac{1}{8} a^{\bar{\mu}} a_{\bar{\mu}} 
+ \frac{1}{12} u^{\bar{\mu}} u^{\bar{\nu}} R_{\bar{\mu}\bar{\nu}}
- \frac{1}{12} \bar{R}, \\ 
\psi^1_{\bar{\alpha}} &= 
-\frac{5}{16} a^{\bar{\mu}} a_{\bar{\mu}} a_{\bar{\alpha}} 
 + \frac{1}{8} \ddot{a}_{\bar{\alpha}} 
+ \frac{1}{8} a_{\bar{\alpha}} u^{\bar{\mu}} u^{\bar{\nu}} 
    R_{\bar{\mu}\bar{\nu}}
- \frac{1}{24} \bar{R} a_{\bar{\alpha}} 
\nonumber \\ & \quad \mbox{}
+ \frac{1}{8} u^{\bar{\mu}} a^{\bar{\nu}} u^{\bar{\lambda}} 
  R_{\bar{\mu}\bar{\nu}\bar{\lambda}\bar{\alpha}} 
+ \frac{1}{12} a^{\bar{\mu}} R_{\bar{\mu}\bar{\alpha}} 
+ \frac{1}{12} u^{\bar{\mu}} u^{\bar{\nu}} 
    R_{\bar{\alpha}\bar{\mu};\bar{\nu}}
\nonumber \\ & \quad \mbox{} 
- \frac{1}{24} u^{\bar{\mu}} u^{\bar{\nu}} 
    R_{\bar{\mu}\bar{\nu};\bar{\alpha}} 
+ \frac{1}{24} R_{;\bar{\alpha}}. 
\end{align} 
\end{subequations} 
The actual expression for
$\psi^0_{\bar{\alpha}\bar{\beta}\bar{\gamma}}$ is obtained from what
appears above by symmetrizing over all three indices; this operation
was suppressed to keep the notation uncluttered.  

Noting that the vector $\sigma^{\bar{\alpha}}$ has a vanishing time
component when $x$ and $\bar{x}$ are simultaneous events, we may
re-express Eq.~(\ref{G3S_local1}) as 
\begin{align} 
G^{\sf S}_3(\bm{x},\bm{\bar{x}}) &= \frac{1}{s} \biggl\{  
1 + \psi^0_{\bar{a}} \sigma^{\bar{a}}
+ \frac{1}{2} \psi^0_{\bar{a}\bar{b}}
   \sigma^{\bar{a}} \sigma^{\bar{b}}
\nonumber \\ & \quad \mbox{} 
+ \frac{1}{6} \psi^0_{\bar{a}\bar{b}\bar{c}} 
   \sigma^{\bar{a}} \sigma^{\bar{b}} \sigma^{\bar{c}}  
+ O(\epsilon^4) 
\nonumber \\ & \quad \mbox{} 
+ s^2 \Bigl[ \psi^1 + \psi^1_{\bar{a}} \sigma^{\bar{a}} 
+ O(\epsilon^2) \Bigr] \biggr\}.  
\label{G3S_local2} 
\end{align} 
And with the results derived in Sec.~\ref{sec:static}, the expansion
coefficients become 
\begin{subequations} 
\label{psi0_coefficients} 
\begin{align} 
\psi^0_{\bar{a}} &= \frac{1}{2} A_{\bar{a}}, \\ 
\psi^0_{\bar{a}\bar{b}} &= -\frac{1}{2} A_{\bar{a}|\bar{b}} 
+ \frac{1}{4} A_{\bar{a}} A_{\bar{b}} 
+ \frac{1}{6} R_{\bar{a}\bar{b}}, \\  
\psi^0_{\bar{a}\bar{b}\bar{c}} &= 
\frac{1}{2} A_{(\bar{a}|\bar{b}\bar{c})}  
- \frac{3}{4} A_{(\bar{a}} A_{\bar{b}|\bar{c})} 
+ \frac{1}{8} A_{\bar{a}} A_{\bar{b}} A_{\bar{c}} 
\nonumber \\ & \quad \mbox{} 
+ \frac{1}{4} A_{(\bar{a}} R_{\bar{b}\bar{c})} 
- \frac{1}{4} R_{(\bar{a}\bar{b}|\bar{c})}. 
\end{align} 
\end{subequations} 
and 
\begin{subequations} 
\label{psi1_coefficients} 
\begin{align} 
\psi^1 &= \frac{1}{4} A^{\bar{a}}_{\ |\bar{a}} 
+ \frac{1}{8} A^{\bar{a}} A_{\bar{a}} - \frac{1}{12} \bar{R}, \\ 
\psi^1_{\bar{a}} &= -\frac{1}{8} A^{\bar{c}}_{\ |\bar{c}\bar{a}}  
- \frac{1}{8} A^{\bar{c}} A_{\bar{c}|\bar{a}} 
+ \frac{1}{8} A^{\bar{c}}_{\ |\bar{c}} A_{\bar{a}} 
+ \frac{1}{16} A^{\bar{c}} A_{\bar{c}} A_{\bar{a}} 
\nonumber \\ & \quad \mbox{} 
- \frac{1}{24} \bar{R} A_{\bar{a}} 
+ \frac{1}{24} R_{|\bar{a}}.   
\end{align} 
\end{subequations} 
Comparing Eq.~(\ref{psi0_coefficients}) with (\ref{W0_coefficients}), 
and Eq.~(\ref{psi1_coefficients}) with (\ref{W1_coefficients}), we
observe that the expansion coefficients of $G^{\sf S}_3$ and 
$G^{\sf H}_3$ are in precise agreement. This allows us to conclude
that 
\begin{equation} 
G^{\sf S}_3(\bm{x},\bm{\bar{x}}) 
= G^{\sf H}_3(\bm{x},\bm{\bar{x}}) + O(\epsilon^3) 
\label{G3S_vs_G3H} 
\end{equation} 
for a static spacetime. 

\subsection{Electromagnetic field} 
\label{subsec:DW_em} 

We next turn to the three-dimensional version of the Detweiler-Whiting
singular Green's function for a static electromagnetic field in a
static spacetime. By virtue of Eq.~(\ref{Phit_static}), we have that
the vector potential of a point charge $e$ situated at $\bm{z}$ is
given by   
\begin{equation} 
\Phi_t^{\sf S}(\bm{x}) = -e N(\bm{z}) G^{\sf S}_3(\bm{x},\bm{z}), 
\label{PhitS_def} 
\end{equation} 
with $G^{\sf S}_3(\bm{x},\bm{z})$ denoting the three-dimensional
version of the Detweiler-Whiting electromagnetic Green's function. And
according to Sec.~18.2 of Ref.~\cite{poisson-pound-vega:11}, we
have that the vector potential is given     
\begin{align} 
\Phi^{\sf S}_\alpha(x) &=  
\frac{e}{2r} U_{\alpha\beta'}(x,x') u^{\beta'}
+ \frac{e}{2r_{\rm adv}} U_{\alpha\beta''}(x,x'') u^{\beta''}  
\nonumber \\ & \quad \mbox{} 
- \frac{e}{2} \int_u^v V_{\alpha\mu}(x,z) u^\mu\, d\tau, 
\label{G3Sem_def2}
\end{align} 
in which $U_{\alpha\mu}(x,z)$, $V_{\alpha\mu}(x,z)$ are the
two-point functions that appear in the construction of the
four-dimensional Green's function.   

To calculate $\Phi^{\sf S}_t$ and obtain $G^{\sf S}_3$ we once more
follow the methods of Haas and Poisson (HP) \cite{haas-poisson:06}, as
outlined in the scalar case. We thus define the world-line functions   
\begin{subequations} 
\begin{align} 
\sigma(\tau) &:= \sigma\bigl(x,z(\tau)\bigr), \\ 
U_\alpha(\tau) &:= U_{\alpha\mu}\bigl(x,z(\tau)\bigr) u^\mu(\tau), \\ 
V_\alpha(\tau) &:= V_{\alpha\mu}\bigl(x,z(\tau)\bigr) u^\mu(\tau),
\end{align} 
\end{subequations} 
in which $x$ is kept fixed. These functions are all scalars with
respect to their dependence upon $z(\tau)$. As in the scalar case they
are expressed as Taylor expansions about $\tau = \bar{\tau}$, at which
$z = \bar{x}$, and the results are converted into explicit expressions
for $r$, $r_{\rm adv}$, $U_{\alpha\beta'} u^{\beta'}$,
$U_{\alpha\beta''} u^{\beta''}$, and the tail integral. The results
for $r$ and $r_{\rm adv}$ appear in Eq.~(\ref{r_Delta}). 

To compute $U_{\alpha\beta'} u^{\beta'}$ and 
$U_{\alpha\beta''} u^{\beta''}$ we write  
\begin{subequations}
\begin{align} 
U_{\alpha\beta'} u^{\beta'} &= U_{\alpha} 
+ \dot{U}_\alpha \Delta_- 
+ \frac{1}{2} \ddot{U}_\alpha \Delta_-^2 
+ \frac{1}{6} \dddot{U}_\alpha \Delta_-^3 
\nonumber \\ & \quad \mbox{} 
+ O(\epsilon^4), \\ 
U_{\alpha\beta''} u^{\beta''} &= U_{\alpha} 
+ \dot{U}_\alpha \Delta_+ 
+ \frac{1}{2} \ddot{U}_\alpha \Delta_+^2 
+ \frac{1}{6} \dddot{U}_\alpha \Delta_+^3 
\nonumber \\ & \quad \mbox{} 
+ O(\epsilon^4),
\end{align} 
\end{subequations} 
in which $U_\alpha(\tau)$ and its derivatives are evaluated at $\tau =
\bar{\tau}$. These quantities are given by 
\begin{widetext} 
\begin{subequations} 
\begin{align} 
U_\alpha &= U_{\alpha\bar{\alpha}}  u^{\bar{\alpha}} 
\nonumber \\
&= g_{\alpha\bar{\alpha}} u^{\bar{\alpha}} \biggl( 1 
+ \frac{1}{12} R_{\sigma\sigma} 
- \frac{1}{24} R_{\sigma\sigma;\sigma} + O(\epsilon^4) \biggr), \\
\dot{U}_\alpha &= U_{\alpha\bar{\alpha};\bar{\beta}} 
u^{\bar{\alpha}} u^{\bar{\beta}} + U_{\alpha\bar{\alpha}}  
a^{\bar{\alpha}} 
\nonumber \\ 
&= g_{\alpha\bar{\alpha}} \biggl[ 
\frac{1}{2} R^{\bar{\alpha}}_{\ uu\sigma}
- \frac{1}{6} R^{\bar{\alpha}}_{\ uu\sigma;\sigma}
+ u^{\bar{\alpha}} \biggl( \frac{1}{6} R_{u\sigma} 
+ \frac{1}{24} R_{\sigma\sigma;u} 
- \frac{1}{12} R_{u\sigma;\sigma} \biggr) 
+ a^{\bar{\alpha}} \biggl( 1 + \frac{1}{12} R_{\sigma\sigma} 
\biggr) + O(\epsilon^3) \biggr], \\ 
\ddot{U}_\alpha &= U_{\alpha\bar{\alpha};\bar{\beta}\bar{\gamma}}  
u^{\bar{\alpha}} u^{\bar{\beta}} u^{\bar{\gamma}} 
+ U_{\alpha\bar{\alpha};\bar{\beta}} \bigl( 
u^{\bar{\alpha}} a^{\bar{\beta}} 
+ 2 a^{\bar{\alpha}} u^{\bar{\beta}} \bigr) 
+ U_{\alpha\bar{\alpha}}  \dot{a}^{\bar{\alpha}} 
\nonumber \\ 
&=  g_{\alpha\bar{\alpha}} \biggl[ 
\frac{1}{3} R^{\bar{\alpha}}_{\ uu\sigma;u}
+ \frac{1}{2} R^{\bar{\alpha}}_{\ ua\sigma}
+ R^{\bar{\alpha}}_{\ au\sigma} 
+ u^{\bar{\alpha}} \biggl( \frac{1}{6} R_{uu} 
+ \frac{1}{6} R_{u\sigma;u} 
- \frac{1}{12} R_{uu;\sigma} 
+ \frac{1}{6} R_{a\sigma} \biggr) 
\nonumber \\ & \quad \mbox{} 
+ a^{\bar{\alpha}} \biggl( \frac{1}{3} R_{u\sigma} \biggr) 
+ \dot{a}^{\bar{\alpha}} + O(\epsilon^2) \biggr],\\ 
\dddot{U}_\alpha &=
U_{\alpha\bar{\alpha};\bar{\beta}\bar{\gamma}\bar{\delta}}    
u^{\bar{\alpha}} u^{\bar{\beta}} u^{\bar{\gamma}} u^{\bar{\delta}}  
+ U_{\alpha\bar{\alpha};\bar{\beta}\bar{\gamma}}  \bigl( 
3 a^{\bar{\alpha}} u^{\bar{\beta}} u^{\bar{\gamma}} 
+ 2 u^{\bar{\alpha}} a^{\bar{\beta}} u^{\bar{\gamma}} 
+ u^{\bar{\alpha}} u^{\bar{\beta}} a^{\bar{\gamma}} \bigr) 
+ U_{\alpha\bar{\alpha};\bar{\beta}} \bigl( 
3 a^{\bar{\alpha}} a^{\bar{\beta}} 
+ 3 \dot{a}^{\bar{\alpha}} u^{\bar{\beta}} 
+ u^{\bar{\alpha}} \dot{a}^{\bar{\beta}} \bigr) 
+ U_{\alpha\bar{\alpha}}  \ddot{a}^{\bar{\alpha}} 
\nonumber \\ 
&=  g_{\alpha\bar{\alpha}} \biggl[ 
\frac{1}{2} R^{\bar{\alpha}}_{\ uau}
+ u^{\bar{\alpha}} \biggl( \frac{1}{2} R_{au} 
+ \frac{1}{4} R_{uu;u} \biggr) 
+ a^{\bar{\alpha}} \biggl( \frac{1}{2} R_{uu} \biggr) 
+ \ddot{a}^{\bar{\alpha}} + O(\epsilon) \biggr]. 
\end{align} 
\end{subequations} 
\end{widetext}
The expansions involve components of the Riemann tensor such as  
$R^{\bar{\alpha}}_{\ uu\sigma} := 
R^{\bar{\alpha}}_{\ \bar{\mu}\bar{\beta}\bar{\nu}}  
u^{\bar{\mu}} u^{\bar{\beta}} \sigma^{\bar{\nu}}$ and components of 
the Ricci tensor such as $R_{\sigma\sigma} :=
R_{\bar{\alpha}\bar{\beta}} \sigma^{\bar{\alpha}}
\sigma^{\bar{\beta}}$. They involve also 
$g^\alpha_{\ \bar{\alpha}} (x,\bar{x})$, the parallel propagator from
$\bar{x}$ to $x$. To arrive at these results we rely on the expansion
of the two-point function $U^\alpha_{\ \bar{\alpha}}(x,\bar{x})$ given
by  
\begin{equation} 
U^\alpha_{\ \bar{\alpha}} = g^\alpha_{\ \bar{\alpha}}\biggl( 
1 + \frac{1}{12} R_{\bar{\mu}\bar{\nu}} \sigma^{\bar{\mu}}
\sigma^{\bar{\nu}} - \frac{1}{24} R_{\bar{\mu}\bar{\nu};\bar{\lambda}}
\sigma^{\bar{\mu}} \sigma^{\bar{\nu}} \sigma^{\bar{\lambda}} 
+ O(\epsilon^3) \biggr); 
\end{equation} 
this is derived, for example, in Appendix B of
Ref.~\cite{anderson-flanagan-ottewill:05}. Another useful expansion is   
\begin{equation} 
g^\alpha_{\ \bar{\alpha};\bar{\beta}} 
= g^\alpha_{\ \bar{\gamma}} \biggl( 
\frac{1}{2} R^{\bar{\gamma}}_{\ \bar{\alpha}\bar{\beta}\bar{\mu}} 
\sigma^{\bar{\mu}}- \frac{1}{6} 
R^{\bar{\gamma}}_{\ \bar{\alpha}\bar{\beta}\bar{\mu};\bar{\nu}}  
\sigma^{\bar{\mu}} \sigma^{\bar{\nu}} + O(\epsilon^3) \biggr).
\end{equation} 
These are differentiated repeatedly with respect to
$\bar{x}^\alpha$, and the results are inserted within the expressions
for $U_\alpha$ and its derivatives.  

To evaluate the tail integral we expand $V_\alpha(\tau)$ as
$V_\alpha(\bar{\tau}) + (\tau-\bar{\tau})\dot{V}_\alpha(\bar{\tau}) 
+ O(\epsilon^2)$ and integrate with respect to $\tau$ between $u =
\bar{\tau} + \Delta_-$ and $v = \bar{\tau} + \Delta_+$. The result is  
\begin{align} 
\int_u^v V_{\alpha\mu} u^\mu\, d\tau 
 &= V_\alpha (\Delta_+ - \Delta_-) 
+ \frac{1}{2} \dot{V}_\alpha (\Delta^2_+ - \Delta^2_-) 
\nonumber \\ & \quad \mbox{} 
+ O(\epsilon^3), 
\end{align} 
in which $V_\alpha = V_{\alpha\bar{\alpha}} u^{\bar{\alpha}}$ and 
$\dot{V}_\alpha = V_{\alpha\bar{\alpha};\bar{\beta}} u^{\bar{\alpha}}
u^{\bar{\beta}} + V_{\alpha\bar{\alpha}} a^{\bar{\alpha}}$. To compute
these quantities we rely on the expansion
\begin{align} 
V^\alpha_{\ \bar{\alpha}} &= g^\alpha_{\ \bar{\gamma}} \biggl[ 
-\frac{1}{2} \biggl( R^{\bar{\gamma}}_{\ \bar{\alpha}} 
- \frac{1}{6} \delta^{\bar{\gamma}}_{\ \bar{\alpha}} \bar{R} \biggr) 
+ \frac{1}{12} 
R^{\bar{\gamma}\ \ \ \ ;\bar{\mu}}_{\ \bar{\alpha}\bar{\mu}\bar{\beta}}  
\sigma^{\bar{\beta}} 
\nonumber \\ & \quad \mbox{} 
+ \frac{1}{4} \biggl( R^{\bar{\gamma}}_{\ \bar{\alpha};\bar{\beta}}  
- \frac{1}{6} \delta^{\bar{\gamma}}_{\ \bar{\alpha}}
\bar{R}_{;\bar{\beta}}  \biggr) \sigma^{\bar{\beta}} 
+ O(\epsilon^2) \biggr], 
\label{Vexp_em} 
\end{align} 
which leads to 
\begin{subequations} 
\begin{align} 
V_\alpha &= g_{\alpha\bar{\alpha}} \biggl[ 
-\frac{1}{2} R^{\bar{\alpha}}_{\ u}
+ \frac{1}{4} R^{\bar{\alpha}}_{\ u;\sigma} 
+ \frac{1}{12} R^{\bar{\alpha}\ \ \ \ ;\bar{\mu}}_{\ u\bar{\mu}\sigma}  
\nonumber \\ & \quad \mbox{} 
+ u^{\bar{\alpha}} \biggl( \frac{1}{12} \bar{R} 
- \frac{1}{24} \bar{R}_{;\sigma}  \biggr) 
+ O(\epsilon^2) \biggr], \\ 
\dot{V}_\alpha &= g_{\alpha\bar{\alpha}} \biggl[ 
-\frac{1}{4} R^{\bar{\alpha}}_{\ u;u}
+ \frac{1}{12} R^{\bar{\alpha}\ \ \ \ ;\bar{\mu}}_{\ u\bar{\mu} u}
- \frac{1}{2} R^{\bar{\alpha}}_{\ a}
\nonumber \\ & \quad \mbox{} 
+ \frac{1}{24} u^{\bar{\alpha}} \bar{R}_{;u} 
+ \frac{1}{12} a^{\bar{\alpha}} \bar{R}   
+ O(\epsilon) \biggr]. 
\end{align} 
\end{subequations} 
Here we make use of the notation $R^{\bar{\alpha}}_{\ a} 
:= R^{\bar{\alpha}}_{\ \bar{\beta}} a^{\bar{\beta}}$, 
$R^{\bar{\alpha}}_{\ u;\sigma} := 
R^{\bar{\alpha}}_{\ \bar{\beta};\bar{\gamma}} u^{\bar{\beta}}
\sigma^{\bar{\gamma}}$, and 
$R^{\bar{\alpha}\ \ \ \ ;\bar{\mu}}_{\ u\bar{\mu} u}  
:= R^{\bar{\alpha}\ \ \ \ ;\bar{\mu}}_{\ \bar{\beta}\bar{\mu} 
\bar{\gamma}} u^{\bar{\beta}} u^{\bar{\gamma}}$. In addition,
$\bar{R}$ is the Ricci scalar evaluated at $\bar{x}$, and 
$\bar{R}_{;u} := \bar{R}_{;\bar{\beta}} u^{\bar{\beta}}$. 

To obtain Eq.~(\ref{Vexp_em}) we rely on standard expansion
techniques. The two-point function is required to satisfy the wave
equation   
\begin{equation} 
\Box V^\alpha_{\ \bar{\alpha}} - R^\alpha_{\ \beta} 
V^\beta_{\ \bar{\alpha}} = 0
\end{equation} 
as well as the light-cone equation
\begin{equation} 
V^\alpha_{\ \bar{\alpha};\beta} \sigma^\beta 
+ \frac{1}{2} (\sigma^\beta_{\ \beta} - 2) V^\alpha_{\ \bar{\alpha}} 
= \frac{1}{2} \bigl( \Box U^\alpha_{\ \bar{\alpha}} 
- R^\alpha_{\ \beta}  U^\beta_{\ \bar{\alpha}} \bigr), 
\end{equation} 
which is evaluated at $\sigma(x,\bar{x}) = 0$. The solution is
expressed as the expansion 
\begin{equation} 
V^\alpha_{\ \bar{\alpha}}(x,\bar{x})  = \sum_{n=0} 
V^\alpha_{n \bar{\alpha}}(x,\bar{x}) \sigma^n, 
\end{equation} 
and the wave equation gives rise to a sequence of equations which
determine $V^\alpha_{n \bar{\alpha}}(x,\bar{x})$ from 
$V^\alpha_{n-1 \bar{\alpha}}$; the light-cone equation determines 
$V^\alpha_{0 \bar{\alpha}}$. Because $\sigma = O(\epsilon^2)$, 
$V^\alpha_{\ \bar{\alpha}} = V^\alpha_{0 \bar{\alpha}}$ to order
$\epsilon$, and this can be obtained by inserting the expansion 
\begin{equation} 
V^\alpha_{0 \bar{\alpha}} = g^\alpha_{\ \bar{\gamma}} \Bigl[ 
A^{\bar{\gamma}}_{\ \bar{\alpha}} + 
A^{\bar{\gamma}}_{\ \bar{\alpha}\bar{\beta}} \sigma^{\bar{\beta}} 
+ O(\epsilon^2) \Bigr] 
\end{equation} 
within the light-cone equation. We use the fact that 
$\sigma^\beta_{\ \beta} = 4 + O(\epsilon^2)$ and to compute 
$\Box U^\alpha_{\ \bar{\alpha}}$ we rely on the expansion 
\begin{equation} 
g^\alpha_{\ \bar{\alpha};\beta} = g^{\alpha}_{\ \bar{\gamma}} 
g^{\bar{\beta}}_{\ \beta}  \biggl[ 
\frac{1}{2} R^{\bar{\gamma}}_{\ \bar{\alpha}\bar{\beta}\bar{\mu}} 
\sigma^{\bar{\mu}} - \frac{1}{3} 
R^{\bar{\gamma}}_{\ \bar{\alpha}\bar{\beta}\bar{\mu};\bar{\nu}}  
\sigma^{\bar{\mu}} \sigma^{\bar{\nu}} + O(\epsilon^3) \biggr];  
\end{equation} 
we eventually arrive at 
\begin{equation} 
\Box U^\alpha_{\ \bar{\alpha}} = g^{\alpha}_{\ \bar{\gamma}} 
\biggl[ \frac{1}{6} \delta^{\bar{\gamma}}_{\ \bar{\alpha}} \bar{R} 
- \frac{1}{6} \delta^{\bar{\gamma}}_{\ \bar{\alpha}}
\bar{R}_{;\bar{\nu}} \sigma^{\bar{\nu}}  
+ \frac{1}{3} R^{\bar{\gamma}\ \ \ \ ;\bar{\mu}}_{\
  \bar{\alpha}\bar{\mu} \bar{\nu}} \sigma^{\bar{\nu}}  
+ O(\epsilon^2) \biggr]. 
\end{equation} 
The end result is Eq.~(\ref{Vexp_em}). 

Putting all the ingredients together, we eventually arrive at the
following expansion for $\Phi_\alpha^{\sf S}(x)$:  
\begin{align} 
\Phi_\alpha^{\sf S}(x) &= 
\frac{e}{s} g_{\alpha}^{\ \bar{\lambda}}(x,\bar{x}) \biggl\{  
\phi^0_{\bar{\lambda}} 
+ \phi^0_{\bar{\lambda}\bar{\alpha}} \sigma^{\bar{\alpha}}
+ \frac{1}{2} \phi^0_{\bar{\lambda}\bar{\alpha}\bar{\beta}}
   \sigma^{\bar{\alpha}} \sigma^{\bar{\beta}}
\nonumber \\ & \quad \mbox{} 
+ \frac{1}{6} \phi^0_{\bar{\lambda}\bar{\alpha}\bar{\beta}\bar{\gamma}}  
   \sigma^{\bar{\alpha}} \sigma^{\bar{\beta}} \sigma^{\bar{\gamma}}  
+ O(\epsilon^4) 
\nonumber \\ & \quad \mbox{} 
+ s^2 \Bigl[ \phi^1_{\bar{\lambda}} 
+ \phi^1_{\bar{\lambda}\bar{\alpha}} \sigma^{\bar{\alpha}} 
+ O(\epsilon^2) \Bigr] \biggr\}, 
\label{em_local1} 
\end{align} 
with 
\begin{widetext} 
\begin{subequations} 
\begin{align} 
\phi^0_{\bar{\lambda}} &= u_{\bar{\lambda}}, \\  
\phi^0_{\bar{\lambda}\bar{\alpha}} &= 
\frac{1}{2} u_{\bar{\lambda}} a_{\bar{\alpha}}, \\  
\phi^0_{\bar{\lambda}\bar{\alpha}\bar{\beta}} &= 
u_{\bar{\lambda}} \biggl( 
\frac{3}{4} a_{\bar{\alpha}} a_{\bar{\beta}} 
+ \frac{1}{6} R_{\bar{\alpha}\bar{\beta}} 
- \frac{1}{3}u^{\bar{\mu}} u^{\bar{\nu}}
   R_{\bar{\mu}\bar{\alpha}\bar{\nu}\bar{\beta}} \biggr), \\
\phi^0_{\bar{\lambda}\bar{\alpha}\bar{\beta}\bar{\gamma}} &=  
u_{\bar{\lambda}} \biggr( 
\frac{15}{8} a_{\bar{\alpha}} a_{\bar{\beta}} a_{\bar{\gamma}} 
- \frac{3}{2} a_{\bar{\alpha}} u^{\bar{\mu}} u^{\bar{\nu}}
   R_{\bar{\mu}\bar{\beta}\bar{\nu}\bar{\gamma}}
+ \frac{1}{4} u^{\bar{\mu}} u^{\bar{\nu}}
   R_{\bar{\mu}\bar{\alpha}\bar{\nu}\bar{\beta};\bar{\gamma}} 
+ \frac{1}{4} a_{\bar{\alpha}} R_{\bar{\beta}\bar{\gamma}} 
- \frac{1}{4} R_{\bar{\alpha}\bar{\beta};\bar{\gamma}} \biggr) 
\end{align} 
\end{subequations} 
and 
\begin{subequations} 
\begin{align} 
\phi^1_{\bar{\lambda}} &= u_{\bar{\lambda}} \biggr( 
-\frac{1}{8} a^{\bar{\mu}} a_{\bar{\mu}} 
+ \frac{1}{12} u^{\bar{\mu}} u^{\bar{\nu}} R_{\bar{\mu}\bar{\nu}}
- \frac{1}{12} \bar{R} \biggr) + \frac{1}{2} \dot{a}_{\bar{\lambda}} 
+ \frac{1}{2} u^{\bar{\mu}} R_{\bar{\lambda}\bar{\mu}}, \\ 
\phi^1_{\bar{\lambda}\bar{\alpha}} &= u_{\bar{\lambda}} \biggr( 
-\frac{5}{16} a^{\bar{\mu}} a_{\bar{\mu}} a_{\bar{\alpha}} 
 + \frac{1}{8} \ddot{a}_{\bar{\alpha}} 
+ \frac{1}{8} a_{\bar{\alpha}} u^{\bar{\mu}} u^{\bar{\nu}} 
    R_{\bar{\mu}\bar{\nu}}
- \frac{1}{24} \bar{R} a_{\bar{\alpha}} 
+ \frac{1}{8} u^{\bar{\mu}} a^{\bar{\nu}} u^{\bar{\rho}} 
  R_{\bar{\mu}\bar{\nu}\bar{\rho}\bar{\alpha}} 
+ \frac{1}{12} a^{\bar{\mu}} R_{\bar{\mu}\bar{\alpha}} 
\nonumber \\ & \quad \mbox{} 
+ \frac{1}{12} u^{\bar{\mu}} u^{\bar{\nu}} 
    R_{\bar{\alpha}\bar{\mu};\bar{\nu}}
- \frac{1}{24} u^{\bar{\mu}} u^{\bar{\nu}} 
    R_{\bar{\mu}\bar{\nu};\bar{\alpha}} 
+ \frac{1}{24} R_{;\bar{\alpha}} \biggr) 
+ a_{\bar{\lambda}} \biggl( \frac{1}{2} \dot{a}_{\bar{\alpha}} 
+ \frac{1}{6} u^{\bar{\mu}} R_{\bar{\alpha}\bar{\mu}} \biggr) 
+ \frac{3}{4} \dot{a}_{\bar{\lambda}} a_{\bar{\alpha}} 
+ \frac{1}{4} u^{\bar{\mu}} a^{\bar{\nu}}
   R_{\bar{\lambda}\bar{\mu}\bar{\nu}\bar{\alpha}}
\nonumber \\ & \quad \mbox{} 
+ \frac{1}{2} u^{\bar{\nu}} a^{\bar{\mu}}
   R_{\bar{\lambda}\bar{\mu}\bar{\nu}\bar{\alpha}}  
+ \frac{1}{6} u^{\bar{\mu}} u^{\bar{\nu}} u^{\bar{\rho}} 
   R_{\bar{\lambda}\bar{\mu}\bar{\nu}\bar{\alpha};\bar{\rho}} 
- \frac{1}{12} u^{\bar{\mu}} \nabla^{\bar{\rho}}
   R_{\bar{\lambda}\bar{\mu}\bar{\rho}\bar{\alpha}}     
+ \frac{1}{4} u^{\bar{\mu}} R_{\bar{\lambda}\bar{\mu}}
   a_{\bar{\alpha}}  
- \frac{1}{4} u^{\bar{\mu}} R_{\bar{\lambda}\bar{\mu};\bar{\alpha}}.  
\end{align} 
\end{subequations} 
\end{widetext}
The actual expression for
$\phi^0_{\bar{\lambda}\bar{\alpha}\bar{\beta}\bar{\gamma}}$ is
obtained from what appears above by symmetrizing over the last three 
indices; this operation was suppressed to keep the notation
uncluttered.    

From Eq.~(\ref{em_local1}) we wish to obtain a more explicit
expression for $\Phi^{\sf S}_t$, and this requires a computation of
the operator of parallel transport. Our considerations near
Eq.~(\ref{parallel_transport}) imply that its components are given by  
\begin{equation} 
g^t_{\ \bar{t}} = \frac{N(\bm{\bar{x}})}{N(\bm{x})}, \qquad 
g^a_{\ \bar{b}} = h^a_{\ \bar{b}}, 
\end{equation} 
in which $h^a_{\ \bar{b}}$ is the operator of parallel transport in
the three-dimensional space; the mixed components $g^t_{\ \bar{a}}$
and $g^a_{\ \bar{t}}$ vanish. Noting that the vector
$\sigma^{\bar{\alpha}}$ has a vanishing time component when $x$ and
$\bar{x}$ are simultaneous events, and making use of the results
derived in Sec.~\ref{sec:static}, we may re-express
Eq.~(\ref{em_local1}) as  
\begin{align} 
\Phi^{\sf S}_t(\bm{x}) &= -\frac{e}{s} N(\bm{x}) \biggl\{  
1 + \phi^0_{\bar{a}} \sigma^{\bar{a}}
+ \frac{1}{2} \phi^0_{\bar{a}\bar{b}}
   \sigma^{\bar{a}} \sigma^{\bar{b}}
\nonumber \\ & \quad \mbox{} 
+ \frac{1}{6} \phi^0_{\bar{a}\bar{b}\bar{c}} 
   \sigma^{\bar{a}} \sigma^{\bar{b}} \sigma^{\bar{c}}  
+ O(\epsilon^4) 
\nonumber \\ & \quad \mbox{} 
+ s^2 \Bigl[ \phi^1 + \phi^1_{\bar{a}} \sigma^{\bar{a}} 
+ O(\epsilon^2) \Bigr] \biggr\},  
\label{em_local2} 
\end{align} 
with
\begin{subequations} 
\label{phi0_coefficients} 
\begin{align} 
\phi^0_{\bar{a}} &= \frac{1}{2} A_{\bar{a}}, \\ 
\phi^0_{\bar{a}\bar{b}} &= -\frac{1}{2} A_{\bar{a}|\bar{b}} 
+ \frac{1}{4} A_{\bar{a}} A_{\bar{b}} 
+ \frac{1}{6} R_{\bar{a}\bar{b}}, \\  
\phi^0_{\bar{a}\bar{b}\bar{c}} &= 
\frac{1}{2} A_{(\bar{a}|\bar{b}\bar{c})}  
- \frac{3}{4} A_{(\bar{a}} A_{\bar{b}|\bar{c})} 
+ \frac{1}{8} A_{\bar{a}} A_{\bar{b}} A_{\bar{c}} 
\nonumber \\ & \quad \mbox{} 
+ \frac{1}{4} A_{(\bar{a}} R_{\bar{b}\bar{c})} 
- \frac{1}{4} R_{(\bar{a}\bar{b}|\bar{c})}. 
\end{align} 
\end{subequations} 
and 
\begin{subequations} 
\label{phi1_coefficients} 
\begin{align} 
\phi^1 &= -\frac{1}{4} A^{\bar{a}}_{\ |\bar{a}} 
+ \frac{1}{8} A^{\bar{a}} A_{\bar{a}} - \frac{1}{12} \bar{R}, \\ 
\phi^1_{\bar{a}} &= \frac{1}{8} A^{\bar{c}}_{\ |\bar{c}\bar{a}}  
- \frac{1}{8} A^{\bar{c}} A_{\bar{c}|\bar{a}} 
- \frac{1}{8} A^{\bar{c}}_{\ |\bar{c}} A_{\bar{a}} 
+ \frac{1}{16} A^{\bar{c}} A_{\bar{c}} A_{\bar{a}} 
\nonumber \\ & \quad \mbox{} 
- \frac{1}{24} \bar{R} A_{\bar{a}} 
+ \frac{1}{24} R_{|\bar{a}}.   
\end{align} 
\end{subequations} 

From Eq.~(\ref{PhitS_def}) and (\ref{em_local2}) we see that the
three-dimensional Green's function involves the ratio
$N(\bm{x})/N(\bm{\bar{x}})$. This can be expressed as an expansion
about $\bm{x} = \bm{\bar{x}}$ by making use of the generalized Taylor
series   
\begin{align} 
N(\bm{x}) &= N(\bm{\bar{x}}) - N_{|\bar{a}} \sigma^{\bar{a}} 
+ \frac{1}{2} N_{|\bar{a}\bar{b}} \sigma^{\bar{a}} \sigma^{\bar{b}} 
- \frac{1}{6} N_{|\bar{a}\bar{b}\bar{c}} \sigma^{\bar{a}}
\sigma^{\bar{b}} \sigma^{\bar{c}} 
\nonumber \\ & \quad \mbox{} 
+ O(\epsilon^4), 
\end{align} 
which leads to 
\begin{align} 
\frac{N(\bm{x})}{N(\bm{\bar{x}})} &= 1 
- A_{\bar{a}} \sigma^{\bar{a}} 
+ \frac{1}{2} \bigl( A_{\bar{a}|\bar{b}} + A_{\bar{a}} A_{\bar{b}}
  \bigr) \sigma^{\bar{a}} \sigma^{\bar{b}} 
\nonumber \\ & \quad \mbox{} 
- \frac{1}{6} \bigl( A_{\bar{a}|\bar{b}\bar{c}} 
+ 3 A_{\bar{a}} A_{\bar{b}|\bar{c}} 
+ A_{\bar{a}} A_{\bar{b}} A_{\bar{c}} \bigr) \sigma^{\bar{a}} 
\sigma^{\bar{b}} \sigma^{\bar{c}} 
\nonumber \\ & \quad \mbox{} 
+ O(\epsilon^4). 
\end{align} 
With this we finally arrive at  
\begin{align} 
G^{\sf S}_3(\bm{x},\bm{\bar{x}}) &= \frac{1}{s} \biggl\{  
1 + \psi^0_{\bar{a}} \sigma^{\bar{a}}
+ \frac{1}{2} \psi^0_{\bar{a}\bar{b}}
   \sigma^{\bar{a}} \sigma^{\bar{b}}
\nonumber \\ & \quad \mbox{} 
+ \frac{1}{6} \psi^0_{\bar{a}\bar{b}\bar{c}} 
   \sigma^{\bar{a}} \sigma^{\bar{b}} \sigma^{\bar{c}}  
+ O(\epsilon^4) 
\nonumber \\ & \quad \mbox{} 
+ s^2 \Bigl[ \psi^1 + \psi^1_{\bar{a}} \sigma^{\bar{a}} 
+ O(\epsilon^2) \Bigr] \biggr\},  
\label{em_local3} 
\end{align}
with
\begin{subequations} 
\label{em0_coefficients} 
\begin{align} 
\psi^0_{\bar{a}} &= -\frac{1}{2} A_{\bar{a}}, \\ 
\psi^0_{\bar{a}\bar{b}} &= \frac{1}{2} A_{\bar{a}|\bar{b}} 
+ \frac{1}{4} A_{\bar{a}} A_{\bar{b}} 
+ \frac{1}{6} R_{\bar{a}\bar{b}}, \\  
\psi^0_{\bar{a}\bar{b}\bar{c}} &= 
-\frac{1}{2} A_{(\bar{a}|\bar{b}\bar{c})}  
- \frac{3}{4} A_{(\bar{a}} A_{\bar{b}|\bar{c})} 
- \frac{1}{8} A_{\bar{a}} A_{\bar{b}} A_{\bar{c}} 
\nonumber \\ & \quad \mbox{} 
- \frac{1}{4} A_{(\bar{a}} R_{\bar{b}\bar{c})} 
- \frac{1}{4} R_{(\bar{a}\bar{b}|\bar{c})}. 
\end{align} 
\end{subequations} 
and 
\begin{subequations} 
\label{em1_coefficients} 
\begin{align} 
\psi^1 &= -\frac{1}{4} A^{\bar{a}}_{\ |\bar{a}} 
+ \frac{1}{8} A^{\bar{a}} A_{\bar{a}} - \frac{1}{12} \bar{R}, \\ 
\psi^1_{\bar{a}} &= \frac{1}{8} A^{\bar{c}}_{\ |\bar{c}\bar{a}}  
- \frac{1}{8} A^{\bar{c}} A_{\bar{c}|\bar{a}} 
+ \frac{1}{8} A^{\bar{c}}_{\ |\bar{c}} A_{\bar{a}} 
- \frac{1}{16} A^{\bar{c}} A_{\bar{c}} A_{\bar{a}} 
\nonumber \\ & \quad \mbox{} 
+ \frac{1}{24} \bar{R} A_{\bar{a}} 
+ \frac{1}{24} R_{|\bar{a}}.   
\end{align} 
\end{subequations} 
We notice that the expansion coefficients can be obtained from
Eqs.~(\ref{psi0_coefficients}) and (\ref{psi1_coefficients}) by making
the replacement $A_a \to -A_a$; this was expected since
Eq.~(\ref{emG3_Poisson}) for the electromagnetic Green's function
differs from Eq.~(\ref{G3_Poisson}) for the scalar Green's function by
the sign of $A^a$.  

A comparison between Eq.~(\ref{em_local3}) and Eq.~(\ref{emG3_local})
allows us to conclude that 
\begin{equation} 
G^{\sf S}_3(\bm{x},\bm{\bar{x}}) 
= G^{\sf H}_3(\bm{x},\bm{\bar{x}}) + O(\epsilon^3) 
\label{emG3S_vs_emG3H} 
\end{equation} 
for a static charge in a static spacetime.  

\section{Equality of $G^{\sf S}_3$ and $G^{\sf H}_3$: A conjecture}  
\label{sec:equality} 

\subsection{Scalar field} 
\label{subsec:equality_scalar} 

The result of Eq.~(\ref{G3S_vs_G3H}) suggests that the equality
between the Hadamard and singular Green's functions might be exact,
holding to all orders in $\epsilon$. We re-express
Eq.~(\ref{G3S_def2}) as 
\begin{equation} 
G^{\sf S}_3(\bm{x},\bm{\bar{x}}) = \frac{1}{s} 
W^{\sf S}(\bm{x},\bm{\bar{x}}) 
\label{WS_def1} 
\end{equation} 
with    
\begin{align} 
W^{\sf S}(\bm{x},\bm{\bar{x}}) &:= \frac{1}{2} \biggl[ 
\frac{s}{r} U(x,x') + \frac{s}{r} U(x,x'') 
\nonumber \\ & \quad \mbox{} 
- s \int_u^v V\bigl(x,z(\tau)\bigr)\, d\tau \biggr], 
\label{WS_def2} 
\end{align} 
and conjecture that $W^{\sf S} = W^{\sf H}$, where $W^{\sf H}$
is the two-point function introduced in Eq.~(\ref{G3_Hadamard}).  We
recall that $s^2 := 2\sigma(x,\bar{x})$ is the squared geodesic
distance between $x$ and the simultaneous event $\bar{x}$, 
$x' := z(u)$ is the retarded point on the (static) world line, 
$x'' := z(v)$ is the advanced point, 
$r := \sigma_{\alpha'} u^{\alpha'}$ is the retarded
distance, $r_{\rm adv} := -\sigma_{\alpha''} u^{\alpha''}$ is the
advanced distance, and $U(x,z)$, $V(x,z)$ are the two-point
functions that appear in the construction of the
four-dimensional Green's function. 

As in Sec.~\ref{subsec:localconf_scalar} above, a 
proof of equality would involve three essential steps. First, the
function $W^{\sf S}$ must be shown to satisfy the same differential
equation as $W^{\sf H}$, as displayed in Eq.~(\ref{W_diffeq}); this
property follows immediately from the fact that $G^{\sf S}_3$ is known
to satisfy Eq.~(\ref{G3_Poisson}), just like $G^{\sf H}_3$. Second,
$W^{\sf S}$ must be shown to satisfy the boundary condition of 
Eq.~(\ref{W_boundary}); this property was established previously and
can be seen directly from Eq.~(\ref{G3S_local2}). Third, the function
$W^{\sf S}$ must be shown to be smooth at $\bm{x} = \bm{\bar{x}}$, by
which we mean that the function must be $C^\infty$ when viewed as a
function of $\bm{x}$ with $\bm{\bar{x}}$ fixed; this property ensures
that $W^{\sf S}$ admits an expansion in powers of $\sigma$ as
displayed in Eq.~(\ref{W_expansion}), which is known to be convergent
and unique. The expansion being unique, smoothness therefore ensures
that $W^{\sf S} = W^{\sf H}$. Evidence that $W^{\sf S}$ is smooth to
order $\epsilon^4$ is provided by Eq.~(\ref{G3S_local2}).   

Of the ingredients involved in the make-up of $W^{\sf S}$, the
two-point functions $U(x,z)$ and $V(x,z)$ are known to be smooth, but 
$s$, $r$, $r_{\rm adv}$, $u$, and $v$ are not. Nevertheless, we
conjecture that the combinations 
\begin{equation} 
s/r, \qquad s/r_{\rm adv}, \qquad s(v-u) 
\label{smooth_combo} 
\end{equation} 
are in fact smooth at $\bm{x} = \bm{\bar{x}}$. The first two are
directly involved in Eq.~(\ref{WS_def2}), and the third one also is
involved by virtue of the mean-value theorem, which allows us to write
the integral as 
\begin{equation} 
s \int_u^v V(x,z)\, d\tau = V(x,x^*) s (v-u),  
\end{equation} 
with $x^* := z(\tau^*)$ ($u < \tau^* < v$) representing a middle point 
on the world line. Establishing that $s/r$, $s/r_{\rm adv}$, and
$s(v-u)$ are smooth is sufficient to prove that $W^{\sf S}$ itself is
smooth. 

Some insight can be gained by examining these quantities in Fermi
normal coordinates $(t,x^a)$ attached to the static world line. With
results collected from Sec.~11 of Ref.~\cite{poisson-pound-vega:11},
we have that  
\begin{subequations} 
\begin{align} 
s &= \sqrt{ \delta_{ab} x^a x^b }, \\ 
r &= s \biggl[ 1 + \frac{1}{2} a_a x^a 
- \frac{1}{8} \bigl( a_a x^a \bigr)^2   
- \frac{1}{8} \dot{a}_t s^2 
\nonumber \\ & \quad \mbox{} 
+ \frac{1}{6} R_{tatb} x^a x^b + O(s^3) \biggr], \\ 
r_{\rm adv} &= r + O(s^4), \\ 
u &= t - s\biggl[ 1 - \frac{1}{2} a_a x^a 
+ \frac{3}{8} \bigl( a_a x^a \bigr)^2   
+ \frac{1}{24} \dot{a}_t s^2 
\nonumber \\ & \quad \mbox{} 
- \frac{1}{6} R_{tatb} x^a x^b + O(s^3) \biggr], \\ 
v &= t + s\biggl[ 1 - \frac{1}{2} a_a x^a 
+ \frac{3}{8} \bigl( a_a x^a \bigr)^2   
+ \frac{1}{24} \dot{a}_t s^2 
\nonumber \\ & \quad \mbox{} 
- \frac{1}{6} R_{tatb} x^a x^b + O(s^3) \biggr], 
\end{align}
\end{subequations} 
in which $a_a$, $\dot{a}_t$, and $R_{tatb}$ respectively represent 
components of the acceleration vector, its covariant derivative, and
the Riemann tensor evaluated on the static world line, at which 
$x^a = 0$; terms involving $\dot{a}_a$ were discarded because these   
components vanish for a static world line in a static spacetime. These
results reveal that $s$, $r$, $r_{\rm adv}$, $u$, and $v$ are indeed
not smooth at $x^a = 0$. But they do show that $r/s$, 
$r_{\rm adv}/s$, and 
\begin{align} 
s(v-u) &= 2s^2 \biggl[ 1 - \frac{1}{2} a_a x^a 
+ \frac{3}{8} \bigl( a_a x^a \bigr)^2   
+ \frac{1}{24} \dot{a}_t s^2 
\nonumber \\ & \quad \mbox{} 
- \frac{1}{6} R_{tatb} x^a x^b + O(s^3) \biggr] 
\end{align} 
are smooth to leading orders in an expansion in powers of $x^a$. 

We now proceed with a sketch of what might constitute a general
proof. The method of proof relies on formal power series, {\it which
are all assumed to converge} in a sufficiently small domain. This
rather strong assumption is the main limitation of our argument, and
the reason why we present it as a conjecture and not a proof. It would   
be desirable to either establish the convergence property, or to
devise an alternative method of proof. This shall be left for future
work.    

We return to Eq.~(\ref{Delta1}) and observe that the odd terms in the
expansion vanish by time-reversal invariance: $\dot{\sigma}$,
$\dddot{\sigma}$, and all other odd derivatives of $\sigma(\tau)$ must
vanish on a static world line in a static spacetime. We recall
that $\sigma(\tau) := \sigma(x,z(\tau))$ with $x$ fixed, and state
that each derivative of $\sigma(\tau)$ is smooth at $\bm{x} =
\bm{\bar{x}}$. Equation (\ref{Delta1}) can therefore be written as 
\begin{subequations}
\label{s2_exp} 
\begin{align}  
s^2 &=\Delta^2 \sum_{n=0}^\infty p_n (\Delta^2)^n, \\ 
p_n &:= \frac{2}{(2n+2)!} \bigl( -\sigma^{(2n+2)} \bigr), 
\end{align}
\end{subequations}
in which a bracketed number attached to $\sigma$ indicates the 
number of differentiations with respect to $\tau$; each expansion
coefficient $p_n$ is smooth at $\bm{x} = \bm{\bar{x}}$.  Time-reversal 
invariance implies that $\Delta_\pm = \pm \sqrt{\Delta^2} 
:= \pm \Delta$, and the expansions of Eq.~(\ref{r_Delta}) can be
expressed as  
\begin{subequations}
\label{r_exp} 
\begin{align}  
r &= r_{\rm adv} = \Delta \sum_{n=0}^\infty q_n (\Delta^2)^n, \\
q_n &:= \frac{1}{(2n+1)!} \bigl( -\sigma^{(2n+2)} \bigr), 
\end{align}
\end{subequations}
with $q_n$ smooth at $\bm{x} = \bm{\bar{x}}$. Combining these
results, we have that 
\begin{equation} 
\frac{r}{s} = \frac{r_{\rm adv}}{s} = 
\frac{  \sum_{n=0}^\infty q_n (\Delta^2)^n } 
{ \sqrt{ \sum_{n=0}^\infty p_n (\Delta^2)^n } }. 
\label{rs_exp1} 
\end{equation} 
As stated previously, each sum in this expression is assumed to
converge for $\Delta^2$ sufficiently small. 

We now wish to reverse the expansion of Eq.~(\ref{s2_exp}). According
to Sec.~3.6.25 of Ref.~\cite{abramowitz-stegun:72}, if 
$y = a x + b x^2 + cx^3 + \cdots$, then 
$x = Ay + By^2 + Cy^3 + \cdots$ with 
$a A = 1$, $a^3 B = -b$, $a^5 C = 2b^2 - ac$, and an algorithm is
known to generate all remaining expansion coefficients. The power
series can thus be reversed when $a \neq 0$. In our case $a = p_0 =
-\ddot{\sigma}$ is indeed nonzero, and $a^{-1}$ is smooth at $\bm{x} =
\bm{\bar{x}}$. The reversed series can then be written as 
\begin{equation} 
\Delta^2 =s^2 \sum_{n=0} a_n (s^2)^n, 
\label{Delta2_ser} 
\end{equation} 
for some coefficients $a_n$ that are known to be smooth at $\bm{x} =
\bm{\bar{x}}$. Because $s^2$ is itself smooth, the assumed convergence
of the sum for sufficiently small $s^2$ ensures that $\Delta^2$ is
smooth at $\bm{x} = \bm{\bar{x}}$. Making the substitution in
Eq.~(\ref{rs_exp1}), we find that $r/s$ and $r_{\rm adv}/s$ can be
expressed as 
\begin{equation} 
\frac{r}{s} = \frac{r_{\rm adv}}{s} = 
\frac{  \sum_{n=0}^\infty b_n (s^2)^n } 
{ \sqrt{ \sum_{n=0}^\infty c_n (s^2)^n } } 
\label{rs_exp2} 
\end{equation} 
for some coefficients $b_n$ and $c_n$ that are smooth at $\bm{x} = 
\bm{\bar{x}}$. This reveals that $r/s$ and $r_{\rm adv}/s$ are 
smooth at $\bm{x} = \bm{\bar{x}}$. We next turn to 
$s(v-u) = 2s\Delta$, which is given by  
\begin{equation} 
s(v-u) = s^2 \sqrt{ {\textstyle \sum_{n=0}^\infty a_n (s^2)^n } }
\end{equation} 
and is also seen to be smooth at $\bm{x} = \bm{\bar{x}}$. 

With the stated assumption on the convergence of formal power series,
we have shown that $r/s$, $r_{\rm adv}/s$, and $s(v-u)$ are all smooth
at $\bm{x} = \bm{\bar{x}}$. This implies that $W^{\sf S}$ is smooth,
and establishes the statement that $G^{\sf S}_3$ and $G^{\sf H}_3$ are
strictly equal.  

\subsection{Electromagnetic field} 
\label{subsec:equality_em} 

The result of Eq.~(\ref{emG3S_vs_emG3H}) suggests that the equality
between the Hadamard and singular Green's functions might also be
exact in the case of the electromagnetic field. A proof of this
statement would involve the same steps as in the scalar case, and the
modifications required for the electromagnetic Green's functions are
too modest to merit a separate discussion. As in the scalar case, 
the essential element is the proof $s/r$, $s/r_{\rm adv}$, and
$s(v-u)$ are all smooth at $\bm{x} = \bm{\bar{x}}$. If this can be 
established, then we can claim immediately that $G^{\sf S}_3$ and 
$G^{\sf H}_3$ are indeed equal to all orders.   

\section{Equality of $G^{\sf S}_3$ and $G^{\sf H}_3$ for
  ultrastatic spacetimes} 
\label{sec:equality_ultrastatic} 

\subsection{Scalar field} 
\label{subsec:equalultra_scalar} 

In this section we return to the theme explored in
Sec.~\ref{sec:equality} and provide a complete proof of equality
between the Hadamard construction $G^{\sf H}_3$ and the
three-dimensional version of the Detweiler-Whiting Green's function
$G^{\sf S}_3$ in the case of ultrastatic spacetimes. These spacetimes
have the property that $N = 1$, so that their metric is 
\begin{equation} 
ds^2 = -dt^2 + h_{ab} dx^a dx^b, 
\end{equation} 
a special case of Eq.~(\ref{metric}). The geometrical quantities
associated with ultrastatic spacetimes can be obtained from the
equations displayed in Sec.~\ref{sec:static} by setting 
$A_a := \partial_a \ln N = 0$. 

The geodesics of ultrastatic spacetimes are described by the equation
$t(\lambda) = t(0) + \dot{t}(0) \lambda$, in which $\lambda$ is an
affine parameter and an overdot indicates differentiation with respect
to $\lambda$, as well as the statement that $x^a(\lambda)$ describes 
geodesics of the spatial metric $h_{ab}$. This implies that the
world function is necessarily given by 
\begin{equation} 
\sigma(x,x') = -\frac{1}{2} (t-t')^2 + \sigma_3(\bm{x},\bm{x'}), 
\end{equation} 
in which $\sigma_3(\bm{x},\bm{x'})$ is the three-dimensional version
of the world function, defined with respect to the spatial metric. 

The simplicity extends to the two-point function $U(x,x')$ that enters
the Detweiler-Whiting construction. We may show, in particular, that
$U$ has no dependence on the time coordinates, so that 
\begin{equation} 
U = U(\bm{x},\bm{x'}). 
\end{equation} 
This statement is a consequence of the defining properties of the
two-point function (see Sec.~14.2 of
Ref.~\cite{poisson-pound-vega:11}), that it must satisfy the 
differential equation  
\begin{equation} 
2 \sigma^\alpha \partial_\alpha U 
+ \bigl( \sigma^\alpha_{\ \alpha} - 4 \bigr)  U = 0
\end{equation} 
in the ultrastatic spacetime, together with the coincidence limit
$U(x',x') = 1$. With the stated properties of the world function, this
becomes 
\begin{equation} 
(t-t') \partial_t U + \sigma^a_3 \partial_a U 
+ \frac{1}{2} \bigl( \sigma^{\ a}_{3\ a} - 3 \bigr)  U = 0.  
\end{equation}
The differential equation can be integrated along any spacetime
geodesic that originates at $x'$. We may, in particular, choose a
time-directed geodesic with no spatial displacement, such that
$t(\lambda) = t' + \lambda$ and $x^a(\lambda) = x^{a'}$. For such a
geodesic we have that $\sigma^a_3 = 0$ and 
$\sigma^{\ a}_{3\ a}  = \sigma^{\ a}_{3\ a}(\bm{x'},\bm{x'}) = 3$, and
the differential equation reduces to $(t-t') \partial_t U = 0$. This
implies that the two-point function cannot depend on $t$, and since
its dependence on $t'$ can only be through the combination $t-t'$, it
cannot depend on $t'$. We have, therefore, established the stated
property.  

The absence of a dependence upon $t$ implies that the two-point
function satisfies the purely spatial differential equation 
\begin{equation} 
\sigma^a_3 \partial_a U 
+ \frac{1}{2} \bigl( \sigma^{\ a}_{3\ a} - 3 \bigr)  U = 0  
\end{equation} 
together with the boundary condition $U(\bm{x'},\bm{x'}) = 1$. These
are precisely the defining relations for the Hadamard function
$W_0(\bm{x},\bm{x'})$, as stated in Eqs.~(\ref{Wn_diffeq}) and
(\ref{W0_boundary}). We conclude, therefore, that  
\begin{equation} 
U(\bm{x},\bm{x'}) = W_0(\bm{x},\bm{x'}) 
\label{U_ultrastatic} 
\end{equation} 
for ultrastatic spacetimes.  
 
Next we turn our attention to the two-point function $V(x,x')$, and
prove that it admits the expansion
\begin{equation} 
V(x,x') = \sum_{n=0}^\infty V_n(\bm{x},\bm{x'}) \sigma^n, 
\label{V_ultrastatic} 
\end{equation} 
in which the coefficients $V_n$ are smooth and time-independent; the
expansion involves the four-dimensional world function, and it is
known to converge within a sufficiently small neighborhood of
$x'$. The proof of the statement relies on the recurrence relations
satisfied by the expansion coefficients \cite{dewitt-brehme:60}, 
\begin{equation} 
\sigma^\alpha \partial_\alpha V_0 
+ \frac{1}{2} \bigl( \sigma^\alpha_{\ \alpha} - 2 \bigr) V_0
= \frac{1}{2} \Box U \biggr|_{\sigma = 0} 
\label{V0_recur} 
\end{equation} 
when $n=0$, and  
\begin{equation} 
\sigma^\alpha \partial_\alpha V_n 
+ \frac{1}{2} \bigl( \sigma^\alpha_{\ \alpha} + 2n - 2 \bigr) V_n 
= -\frac{1}{2n} \Box V_{n-1} 
\label{Vn_recur} 
\end{equation} 
when $n > 0$. 

We begin with an examination of $V_0$. In ultrastatic spacetimes its
differential equation becomes 
\begin{equation} 
(t-t') \partial_t V_0 + \sigma^a_{3} \partial_a V_0 
+ \frac{1}{2} \bigl( \sigma^{\ a}_{3\ a} - 1 \bigr) V_0 
= \frac{1}{2} \nabla^2 U. 
\end{equation} 
Once more this equation can be integrated along any spacetime geodesic
that originates at $x'$, and once more we choose a time-directed
geodesic. In this case we have 
\begin{equation} 
(t-t') \partial_t V_0 + V_0 
= \frac{1}{2} \nabla^2 U \biggr|_{\bm{x}=\bm{x'}} 
= \frac{1}{12} R(\bm{x'}), 
\end{equation} 
in which $R(\bm{x'})$ is the spatial Ricci scalar evaluated at
$\bm{x'}$, obtained from the known expression for $\nabla^2 U$
evaluated in the coincidence limit (see Sec.~14.2 of
Ref.~\cite{poisson-pound-vega:11}). The general solution to this
equation is $V_0 = \frac{1}{12} R(\bm{x'}) + c(t-t')^{-1}$ where $c$ 
is a constant, and we see that $V_0$ fails to be smooth at $x=x'$
unless $c = 0$. We conclude that $V_0$ cannot depend on time. 

Turning next to $V_n$, we proceed by induction. We assume that
$V_{n-1}$ is known to be time-independent, and prove that $V_n$
must in turn be time-independent. We begin with the differential
equation 
\begin{equation} 
(t-t') \partial_t V_n 
+ \sigma_3^a \partial_a V_n 
+ \frac{1}{2} \bigl( \sigma^{\ a}_{3\ a} + 2n - 1 \bigr) V_n 
= -\frac{1}{2n} \nabla^2 V_{n-1},  
\end{equation} 
which we integrate along a time-directed geodesic. The equation
reduces to 
\begin{equation} 
(t-t') \partial_t V_n 
+ (n+1) V_n = -\frac{1}{2n} \nabla^2 V_{n-1} \biggr|_{\bm{x}=\bm{x'}},  
\end{equation} 
and we find that the general solution contains a term 
$c (t-t')^{-(n+1)}$ that fails to be smooth at $x=x'$ unless 
$c = 0$. This allows us to conclude that $V_n$ cannot depend on time,
and we have established Eq.~(\ref{V_ultrastatic}).   

We may now demonstrate the equality of the Green's functions. The 
four-dimensional version of the Detweiler-Whiting singular Green's
function is 
\begin{equation} 
G^{\sf S}_4(x,x') = \frac{1}{2} U(x,x') \delta(\sigma) 
- \frac{1}{2} V(x,x') \Theta (\sigma), 
\end{equation} 
in which $\Theta$ is the Heaviside step function and $\delta$ the
Dirac distribution. According to Eq.~(\ref{G3_and_G4}), the
three-dimensional version is 
\begin{equation} 
G^{\sf S}_3(\bm{x},\bm{x'})= \int G_4^{\sf S}(x,x')\, dt'
\end{equation} 
when $N(\bm{x'}) = 1$. With $U$ independent of time and $\sigma$
factorized as   
\begin{equation}  
\sigma = -\frac{1}{2} \Bigl( \Delta t - \sqrt{2\sigma_3} \Bigr) 
\Bigl( \Delta t + \sqrt{2\sigma_3} \Bigr) 
\end{equation} 
with $\Delta t = t-t'$, we find that the integral becomes 
\begin{equation} 
G^{\sf S}_3(\bm{x},\bm{x'}) = \frac{U(\bm{x},\bm{x'})}{\sqrt{2\sigma_3}} 
- \frac{1}{2} \int_{-\sqrt{\sigma_3}}^{\sqrt{\sigma_3}}  
V(x,x')\, d\Delta t. 
\end{equation} 
In this we insert Eq.~(\ref{V_ultrastatic}), integrate term by term
using 
\begin{align} 
\int_{-\sqrt{\sigma_3}}^{\sqrt{\sigma_3}}  
\sigma^n\, d\Delta t &= 
\biggl( -\frac{1}{2} \biggr)^n 
\int_{-\sqrt{\sigma_3}}^{\sqrt{\sigma_3}}  
\bigl( \Delta t^2 - 2\sigma_3 \bigr)^n \, d\Delta t 
\nonumber \\
&= \frac{\sqrt{\pi} \Gamma(n+1)}{2^n \Gamma(n + \frac{3}{2})} 
(2 \sigma_3)^{n+\frac{1}{2}},  
\end{align} 
and simplify. Our final expression for the singular Green's function
is 
\begin{equation} 
G^{\sf S}_3(\bm{x},\bm{x'}) 
= \frac{W^{\sf S}(\bm{x},\bm{x'})}{\sqrt{2\sigma_3}} 
\end{equation} 
with 
\begin{align} 
W^{\sf S}(\bm{x},\bm{x'}) &= U(\bm{x},\bm{x'}) 
\nonumber \\ & \quad \mbox{}
- \sum_{n=1}^\infty \frac{(n-1)!}{(2n-1)!!} V_{n-1}(\bm{x},\bm{x'}) 
(2 \sigma_3 )^n. 
\end{align} 
These equations reveal that $G^{\sf S}_3$ does admit a
three-dimensional Hadamard form, and that we may make the
identifications  
\begin{equation} 
W^{\sf S}_0(\bm{x},\bm{x'}) := U (\bm{x},\bm{x'}) 
\end{equation} 
as in Eq.~(\ref{U_ultrastatic}), and 
\begin{equation} 
W^{\sf S}_n(\bm{x},\bm{x'}) := -\frac{(n-1)!}{(2n-1)!!} 
V_{n-1}(\bm{x},\bm{x'}). 
\end{equation} 
These coefficients satisfy the recursion relation of
Eq.~(\ref{Wn_diffeq}), as can be seen by invoking 
Eqs.~(\ref{V0_recur}) and (\ref{Vn_recur}), and are therefore the same 
coefficients that appear in Eq.~(\ref{W_expansion}). The proof of
equality between $G^{\sf S}_3(\bm{x},\bm{x'})$ and 
$G^{\sf H}_3(\bm{x},\bm{x'})$ in ultrastatic spacetimes is complete,
and the calculations have revealed the relationship between $U$ and
$W_0$, and between $V_n$ and $W_n$.    

\subsection{Electromagnetic field} 
\label{subsec:equalultra_em} 

The proof of equality between the Hadamard construction $G^{\sf H}_3$
and the three-dimensional version of the Detweiler-Whiting Green's
function $G^{\sf S}_3$ for ultrastatic spacetimes proceeds along the
same lines as in the scalar case. In fact, the calculational details
are strictly identical, because the two-point functions 
$U_t^{\, t'}(x,x')$ and $V_t^{\, t'}(x,x')$ that are involved in the
relevant component of the electromagnetic Green's function, 
\begin{equation} 
G^{{\sf S}\, t'}_{\ t}(x,x') = \frac{1}{2} U_t^{\, t'}(x,x')
\delta(\sigma)  - \frac{1}{2} V_t^{\, t'}(x,x') \Theta (\sigma),  
\end{equation} 
are strictly identical to their scalar counterparts: $U_t^{\, t'} = U$
and $V_t^{\, t'} = V$. The first equality follows from the general
relation $U_\alpha^{\ \beta'} = g_\alpha^{\ \beta'} U$ --- see
Eqs.~(14.8) and (15.9) in Ref.~\cite{poisson-pound-vega:11} ---   
together with the property that $g_t^{\, t'} = 1$ for ultrastatic
spacetimes. The second equality follows from the fact that if
$V_\alpha^{\, \beta'}$ is expanded as 
\begin{equation} 
V_\alpha^{\, \beta'} = \sum_{n=0} V_{n\, \alpha}^{\ \ \beta'} \sigma^n, 
\end{equation} 
then the recursion relations satisfied by $V_{n\, t}^{\ \ t'}$ are
strictly identical to those satisfied by $V_n$. The results of
Sec.~\ref{subsec:equalultra_scalar}, therefore, allow us to
state that for ultrastatic spacetimes, $G^{\sf H}_3 = G^{\sf S}_3$ in 
the electromagnetic case also.   

\section{Self-force in spherical spacetimes} 
\label{sec:spherical} 

\subsection{Scalar field} 
\label{subsec:spherical_scalar} 

We consider the self-force acting on a static scalar charge $q$ in a
static and spherically-symmetric spacetime. The metric is written as  
\begin{equation} 
ds^2 = -e^{2\psi}\, dt^2 + f^{-1} dr^2 
+ r^2 \bigl( d\theta^2 + \sin^2\theta\, d\phi^2 \bigr), 
\label{spherical_metric}
\end{equation} 
in which $\psi$ and $f$ are functions of $r$. In this notation 
$N = e^{\psi}$ and $A_r = \psi'$ is the only nonvanishing component of
the vector $A_a$. The potential $\Phi$ generated by the point scalar
charge is a solution to
\begin{equation} 
\nabla^2 \Phi + A^a \partial_a \Phi = -4\pi q \delta_3(\bm{x},\bm{z}), 
\end{equation} 
which is obtained from Eqs.~(\ref{Phi_Poisson}) and
(\ref{mu_static}). 

In order to integrate the field equation we decompose the potential
and source in spherical harmonics: 
\begin{equation} 
\Phi(t,\theta,\phi) = \sum_{\ell m} \Phi_{\ell m}(r) 
Y_{\ell m}(\theta,\phi) 
\end{equation} 
and 
\begin{equation} 
\delta_3(\bm{x},\bm{z}) = \frac{f_0^{1/2}}{r_0^2} \delta(r-r_0) 
\sum_{\ell m} Y^*_{\ell m}(\theta_0,\phi_0) Y_{\ell m}(\theta,\phi),  
\end{equation} 
in which $(r_0,\theta_0,\phi_0)$ represent the spherical coordinates
of the particle's position $\bm{z}$, and $f_0 := f(r_0)$. Without loss
of generality we may place the particle along the polar axis 
($\theta_0 = 0$) and exploit the property 
$Y_{\ell m}(0,\phi) = \sqrt{(2\ell+1)/(4\pi)} \delta_{m,0}$ of
spherical-harmonic functions. Substitution within the field equation
then produces 
\begin{align} 
r^2 \Phi''_{\ell 0} &+ \biggl( 2 + \frac{rf'}{2f} + r \psi' \biggr) 
r \Phi'_{\ell 0} - \frac{\ell(\ell+1)}{f} \Phi_{\ell 0}   
\nonumber \\ & \quad =
-4\pi q \sqrt{\frac{2\ell + 1}{4\pi}} f_0^{-1/2}\, \delta(r-r_0),  
\label{scalar_l0} 
\end{align} 
in which a prime indicates differentiation with respect to $r$. 
The modes with $m \neq 0$ necessarily vanish. 

The self-force acting on the scalar charge is given by $F^\alpha = q
(g^{\alpha\beta} + u^\alpha u^\beta) \nabla_\beta \Phi^{\sf R}$, in
which $\Phi^{\sf R} := \Phi - \Phi^{\sf S}$ is the difference between
the actual potential $\Phi$ and the Detweiler-Whiting singular field 
$\Phi^{\sf S}$; the regular potential is known to be smooth at 
$\bm{x} = \bm{z}$. In a static situation the self-force has a
vanishing time component, and its spatial components are given by 
$F^a = q h^{ab} \partial_b \Phi^{\sf R}$. In a spherically-symmetric
spacetime the angular components vanish, and the radial component is  
\begin{equation} 
F^r = q f_0 \partial_r \Phi^{\sf R}(r_0,\theta_0,\phi_0). 
\label{self_force1} 
\end{equation} 
Recalling the spherical-harmonic decomposition of the potential, we
may express this as 
\begin{equation} 
F^r = q f_0 \lim_{\bm{x} \to \bm{z}} \sum_{\ell} 
\Bigl[ \bigl( \partial_r \Phi \bigr)_\ell 
- \bigl( \partial_r \Phi^{\sf S} \bigr)_\ell \Bigr], 
\label{self_force2} 
\end{equation} 
in which 
\begin{equation} 
\bigl( \partial_r \Phi \bigr)_\ell := 
\sum_{m=-\ell}^\ell \Phi'_{\ell m}(r) Y_{\ell m}(\theta,\phi) 
\end{equation} 
are the {\it multipole coefficients} of $\partial_r \Phi$, while
$( \partial_r \Phi^{\sf S} )_\ell$ are those of the
singular potential $\Phi^{\sf S}$. Recalling the relation of
Eq.~(\ref{Phi_static}) between the potential and the scalar Green's
function, we may write this in the form 
\begin{equation} 
F^r = q^2 f_0 \lim_{\bm{x} \to \bm{z}} \sum_{\ell} 
\Bigl[ q^{-1} \bigl( \partial_r \Phi \bigr)_\ell
- \bigl( \partial_r G^{\sf S}_3 \bigr)_\ell \Bigr], 
\label{self_force3} 
\end{equation} 
in which $G^{\sf S}_3(\bm{x},\bm{z})$ is the three-dimensional version
of the Detweiler-Whiting singular Green's function introduced in
Sec.~\ref{subsec:DW_scalar}.   

The limit in Eq.~(\ref{self_force3}) can be taken by setting $r = r_0
+ \Delta$, $\theta = \theta_0$, $\phi = \phi_0$, and letting
$\Delta \to 0$ (from either direction). With this choice, we shall
show below that 
\begin{align} 
\bigl( \partial_r G^{\sf S}_3 \bigr)_\ell &= 
A \bigl( \ell + {\textstyle \frac{1}{2}} \bigr) 
+ B + \frac{C}{\bigl( \ell + {\textstyle \frac{1}{2}} \bigr)} 
+ \frac{D}{\bigl( \ell - {\textstyle \frac{1}{2}} \bigr)
    \bigl( \ell + {\textstyle \frac{3}{2}} \bigr)} 
\nonumber \\ & \quad 
+ O(\ell^{-3}), 
\label{reg_param1} 
\end{align} 
in which the {\it regularization parameters} $A$, $B$, $C$, and $D$
depend on $r_0$ but are independent of $\ell$; explicit expressions
will be presented below. Inserting Eq.~(\ref{reg_param1}) within
Eq.~(\ref{self_force3}) provides a practical method of computing the
self-force by means of a regularized mode sum that converges to the
correct answer. With the particle placed on the polar axis 
($\theta_0 = 0$), the multipole coefficients reduce to 
\begin{equation} 
\bigl( \partial_r \Phi \bigr)_\ell 
=  \sqrt{\frac{2\ell + 1}{4\pi}} \Phi'_{\ell 0}(r_0 + \Delta). 
\end{equation} 

To establish the relation of Eq.~(\ref{reg_param1}) and calculate the
regularization parameters we follow the method described in Sec.~V of
Haas and Poisson (HP) \cite{haas-poisson:06}, which we adapt to the
situation at hand. In HP the motion of the particle was geodesic and
the spacetime was that of a Schwarzschild black hole; here the
particle is kept in place in any static, spherically-symmetric
spacetime. In HP the motion was taking place in the equatorial plane,
and a transformation of the angular coordinates was implemented to put
the particle momentarily on the polar axis; here the particle is kept
on the axis at all times, and the transformation is not
required. Following Sec.~III of HP, the singular Green's function of
Eq.~(\ref{G3S_local2}) is expressed as an expansion in powers of the 
coordinate displacements $w^a := x^a - \bar{x}^a$, in which
$\bm{\bar{x}} := \bm{z}$ denotes the particle's position. As in Sec.~V
of HP we express the angular separations $w^\theta$ and $w^\phi$ in
terms of functions $Q := \sqrt{1-\cos\theta}$, $\sin\phi$, 
and $\cos\phi$ that are globally well-defined on the sphere. In this
case of static motion, the squared-distance function introduced in
Eq.~(5.22) of HP reduces to 
\begin{equation} 
\rho^2 = f_0^{-1} \Delta^2 + 2r_0^2 Q 
= 2r_0^2 \bigl( \delta^2 + 1 - \cos\theta \bigr), 
\end{equation} 
where 
\begin{equation} 
\delta^2 = \frac{\Delta^2}{2r_0^2 f_0} 
\end{equation} 
with $\Delta := w^r = r - r_0$. Following the steps outlined in
Sec.~V\,E of HP, we obtain an expansion for $\partial_r G^{\sf S}_3$
that takes the schematic form of 
\begin{align} 
\partial_r G^{\sf S}_3 &= \bigl(  \partial_r G^{\sf S}_3 \bigr)_{-2}
+ \bigl(  \partial_r G^{\sf S}_3 \bigr)_{-1}  
+ \bigl(  \partial_r G^{\sf S}_3 \bigr)_{0}  
+ \bigl(  \partial_r G^{\sf S}_3 \bigr)_{1} 
\nonumber \\ & \quad \mbox{} 
+ O(\epsilon^2), 
\end{align} 
in which a subscript attached to enclosing brackets indicates
the scaling with powers of $\epsilon$. The various terms are
schematically given by 
\begin{subequations} 
\begin{align} 
\bigl(  \partial_r G^{\sf S}_3 \bigr)_{-2} &=
M_{-2} (\Delta/\rho^3), \\
\bigl(  \partial_r G^{\sf S}_3 \bigr)_{-1} &=
M_{-1} (1/\rho) + O(\Delta^2/\rho^3) 
\nonumber \\ & \quad \mbox{} 
+ O(\Delta^4/\rho^5), \\ 
\bigl(  \partial_r G^{\sf S}_3 \bigr)_{0} &= O(\Delta/\rho)  
+ O(\Delta^3/\rho^3) + O(\Delta^5/\rho^5) 
\nonumber \\ & \quad \mbox{} 
+ O(\Delta^7/\rho^7), \\ 
\bigl(  \partial_r G^{\sf S}_3 \bigr)_{1} &= M_{1} \rho 
+ O(\Delta^2/\rho)  + O(\Delta^4/\rho^3) 
+ O(\Delta^6/\rho^5) 
\nonumber \\ & \quad \mbox{} 
+ O(\Delta^8/\rho^7) 
+ O(\Delta^{10}/\rho^9).
\end{align} 
\end{subequations} 
The terms involving the coefficients $M_{-2}$, $M_{-1}$, and
$M_{1}$ are those giving rise to the regularization parameters;
all other terms are unimportant. 

The multipole decomposition of $\partial_r G^{\sf S}_3$ is next
carried out with the help of Eq.~(A19) of Haas and Poisson; because
the expressions are all $\phi$-independent (by virtue of the axial
symmetry of the problem), there is no need to perform the
$\phi$-average described by Eq.~(A13). We make use of the relations  
\begin{subequations} 
\begin{align} 
(\Delta/\rho^3)_\ell &= 
\bigl(\ell+{\textstyle \frac{1}{2}} \bigr) \frac{f^{1/2}_0}{r_0^2}
\,\mbox{sign}(\Delta)+ O(\Delta), \\ 
(1/\rho)_\ell &= \frac{1}{r_0}+ O(\Delta), \\ 
(\rho)_\ell &= -\frac{r_0}{\bigl(\ell - {\textstyle \frac{1}{2}}\bigr) 
  \bigl(\ell + {\textstyle \frac{3}{2}} \bigr)} + O(\Delta)   
\end{align} 
\end{subequations} 
and arrive at the expression of Eq.~(\ref{reg_param1}) with 
$A = M_{-2} f_0^{1/2} r_0^{-2} \mbox{sign}(\Delta)$, 
$B = M_{-1}/r_0$, $C = 0$, and $D = -M_1 r_0$. The detailed
computation reveals that  
\begin{subequations} 
\label{reg_param2} 
\begin{align} 
A &= -\frac{1}{r^2} f^{-1/2}\,\mbox{sign}(\Delta), \\ 
B &= -\frac{1}{2r^2} \bigl( 1 + r \psi' \bigr), \\ 
C &= 0, \\ 
D &= -\frac{1}{16 r^2} \Bigl[ \bigl(1 + r \psi' \bigr) 
- \bigl( 1 + r\psi' + 3 r^2 \psi^{\prime 2} + r^3 \psi^{\prime 3} 
\nonumber \\ & \quad \mbox{} 
- 6 r^2 \psi'' - 2r^3 \psi''' \bigr) f 
+ \bigl( 1 + 4r \psi' + 3r^2 \psi'' \bigr) r f' 
\nonumber \\ & \quad \mbox{} 
+ \bigl( 1 + r\psi' \bigr) r^2 f'' \Bigr], 
\end{align} 
\end{subequations} 
in which all functions are to be evaluated at $r=r_0$. These are the
regularization parameters for a static scalar charge in any static,
spherically-symmetric spacetime. 

\subsection{Electromagnetic field} 
\label{subsec:spherical_em} 

We next consider the self-force acting on a static electric charge $e$
in a static and spherically-symmetric spacetime with the metric of
Eq.~(\ref{spherical_metric}). The vector potential $\Phi_t$ generated
by the point charge is a solution to
\begin{equation} 
\nabla^2 \Phi_t - A^a \partial_a \Phi_t = 
4\pi e N(\bm{z}) \delta_3(\bm{x},\bm{z}),
\end{equation} 
which is obtained from Eqs.~(\ref{Phit_Poisson}) and
(\ref{emmu_static}). 

As in the scalar case we decompose the potential and source in
spherical harmonics and place the particle along the polar axis
($\theta_0 = 0$). Substitution within the field equation then produces  
\begin{align} 
r^2 \Phi''_{t\, \ell 0} &+ \biggl( 2 + \frac{rf'}{2f} - r \psi' \biggr) 
r \Phi'_{t\, \ell 0} - \frac{\ell(\ell+1)}{f} \Phi_{t\, \ell 0}
\nonumber \\ & \quad =
4\pi e \sqrt{\frac{2\ell + 1}{4\pi}} e^{\psi_0} f_0^{-1/2}\,
\delta(r-r_0),   
\label{em_l0}
\end{align} 
in which $f_0 := f(r_0)$, $\psi_0 := \psi(r_0)$, and a prime indicates
differentiation with respect to $r$. The modes with $m \neq 0$
necessarily vanish.  

The self-force acting on the scalar charge is given by 
$F^\alpha = e F^{\ \alpha}_{{\sf R}\ \beta} u^\beta$, in which 
$F^{\ \alpha}_{{\sf R}\ \beta}  := F^{\alpha}_{\ \beta}  
- F^{\ \alpha}_{{\sf S}\ \beta}$ is the difference between
the actual electromagnetic field and the Detweiler-Whiting singular
field; the regular field is known to be smooth at 
$\bm{x} = \bm{z}$. In a static situation the self-force has a
vanishing time component, and in spherical symmetry its radial
component is   
\begin{equation} 
F^r = e e^{-\psi_0} f_0 \partial_r \Phi_t^{\sf R}(r_0,\theta_0,\phi_0). 
\label{emself_force1} 
\end{equation} 
We express this as 
\begin{equation} 
F^r = e e^{-\psi_0} f_0 \lim_{\bm{x} \to \bm{z}} \sum_{\ell}  
\Bigl[ \bigl( \partial_r \Phi_t \bigr)_\ell 
- \bigl( \partial_r \Phi_t^{\sf S} \bigr)_\ell \Bigr], 
\label{emself_force2} 
\end{equation} 
in which 
\begin{align} 
\bigl( \partial_r \Phi_t \bigr)_\ell &:= 
\sum_{m=-\ell}^\ell \Phi'_{t\, \ell m}(r) Y_{\ell m}(\theta,\phi) 
\nonumber \\ 
&= \sqrt{\frac{2\ell + 1}{4\pi}} \Phi'_{t\, \ell 0}(r_0 + \Delta)  
\end{align} 
are the {\it multipole coefficients} of $\partial_r \Phi_t$, while
$( \partial_r \Phi_t^{\sf S} )_\ell$ are those of the
singular potential. Recalling the relation of Eq.~(\ref{Phit_static})
between the potential and the scalar Green's function, we may write
this in the form  
\begin{equation} 
F^r = e^2 f_0 \lim_{\bm{x} \to \bm{z}} \sum_{\ell} 
\Bigl[ e^{-1} e^{-\psi_0} \bigl( \partial_r \Phi_t \bigr)_\ell
+ \bigl( \partial_r G^{\sf S}_3 \bigr)_\ell \Bigr], 
\label{emself_force3} 
\end{equation} 
in which $G^{\sf S}_3(\bm{x},\bm{z})$ is the three-dimensional version
of the Detweiler-Whiting singular Green's function introduced in
Sec.~\ref{subsec:DW_em}.   

The limit in Eq.~(\ref{self_force3}) is taken by setting $r = r_0
+ \Delta$, $\theta = \theta_0 = 0$, and $\phi = \phi_0 = 0$ and
letting $\Delta \to 0$ (from either direction). With this choice, the
multipole coefficients of the singular Green's function take the same
form as in Eq.~(\ref{reg_param1}). In this case, however, because of
the different sign in front of $A^a$ in the Poisson equation for 
$\Phi_t$, the regularization parameters are given by 
\begin{subequations} 
\label{emreg_param} 
\begin{align} 
A &= -\frac{1}{r^2} f^{-1/2}\,\mbox{sign}(\Delta), \\ 
B &= -\frac{1}{2r^2} \bigl( 1 - r \psi' \bigr), \\ 
C &= 0, \\ 
D &= -\frac{1}{16 r^2} \Bigl[ \bigl(1 - r \psi' \bigr) 
- \bigl( 1 - r\psi' + 3 r^2 \psi^{\prime 2} - r^3 \psi^{\prime 3} 
\nonumber \\ & \quad \mbox{}
+ 6 r^2 \psi'' + 2r^3 \psi''' \bigr) f 
+ \bigl( 1 - 4r \psi' - 3r^2 \psi'' \bigr) r f' 
\nonumber \\ & \quad \mbox{}
+ \bigl( 1 - r\psi' \bigr) r^2 f'' \Bigr], 
\end{align} 
\end{subequations} 
in which all functions are to be evaluated at $r=r_0$. These are the
regularization parameters for a static electric charge in any static, 
spherically-symmetric spacetime. The computations that lead to
Eq.~(\ref{emreg_param}) involve the same steps as those described in
Sec.~\ref{subsec:spherical_scalar}. 

\begin{acknowledgments} 
This work was supported by the Natural Sciences and Engineering
Research Council of Canada. M.C.\ acknowledges funding support from a
IRCSET-Marie Curie International Mobility Fellowship in Science,
Engineering and Technology. 
\end{acknowledgments}    

\bibliography{../bib/master} 

\end{document}